# Interpreting the Ionization Sequence in AGN Emission-Line Spectra

Chris T. Richardson,[1,2*] James T. Allen,[3*] Jack A. Baldwin,[1*] Paul C. Hewett[4*] and Gary J. Ferland[5*]

[1]*Physics & Astronomy Department, Michigan State University, East Lansing, MI 48864-1116, USA*
[2]*Physics Department, Elon University, Elon, NC, 27244, USA*
[3]*Sydney Institute for Astronomy, School of Physics, University of Sydney, NSW 2006, Australia*
[4]*Institute of Astronomy, University of Cambridge, Madingley Road, Cambridge, CB3 0HA, UK*
[5]*Physics and Astronomy Department, University of Kentucky, Lexington, KY, 40506-0055, USA*

**Abstract**

We investigate the physical cause of the great range in the ionization level seen in the spectra of narrow lined active galactic nuclei (AGN). We used a recently developed technique called mean field independent component analysis to identify examples of individual SDSS galaxies whose spectra are not dominated by emission due to star formation (SF), which we therefore designate as AGN. We assembled high S/N ratio composite spectra of a sequence of these AGN defined by the ionization level of their narrow-line regions (NLR), and extending down to very low-ionization cases. We then used a local optimally emitting cloud (LOC) model to fit emission-line ratios in this AGN sequence, including the weak lines that can be measured only in the co-added spectra. These weak line ratios provide consistency checks on the density, temperature, abundances and ionizing continuum of Seyfert galaxies determined from strong-line ratios. After integrating over a wide range of clouds at different radii and densities, our models indicate that the radial extent of the NLR is the major parameter in determining the position of higher to moderate ionization AGN along our sequence. This provides a physical interpretation for their systematic variation. Higher ionization AGN contain optimally emitting clouds that are more concentrated towards the central continuum source than in lower ionization AGN. Our LOC models indicate that for the special set of objects that lie on our AGN sequence, the ionizing luminosity is anti-correlated with the NLR ionization level, and hence anticorrelated with the radial concentration and actual physical extent of the NLR. A possible interpretation that deserves further exploration is that the ionization sequence might be an age sequence where low ionization objects are older and have systematically cleared out their central regions by radiation pressure. We consider the alternative that our AGN sequence instead represents a mixing curve combining SF and AGN spectra in different proportions, but argue that while many galaxies in fact do have this type of composite spectra, our AGN sequence appears to be a special set of objects with negligible SF excitation.

---

\* richa684@msu.edu (CTR), j.allen@physics.usyd.edu.au (JTA), baldwin@pa.msu.edu (JAB), phewett@ast.cam.ac.uk (PCH); gary@pa.uky.edu (GJF)

# 1. Introduction

This paper explores the nature of the narrow line region (NLR) in active galactic nuclei (AGN) by comparing plasma simulations to the results obtained from a relatively new and novel technique, mean field independent component analysis (MFICA). This introduction first describes the present understanding of the NLR, follow by the sequence of AGN galaxies defined by MFICA, and finally the goals of this paper.

## 1.1 The Narrow Line Region

The NLR of AGN is a region of ionized and neutral gas generally classified by strong [N II] λ6584[1] and [O III] λ5007 emission. In contrast to the very small broad line region (BLR), the NLR is on the order of $10^3$ pc in size and contains relatively low-density gas, with electron densities $n_e \sim 10^4$ cm$^{-3}$. However, the physical picture of typical NLRs remains unclear. An important route for advancing our understanding of this region is to compare observed emission-line strengths to the ones predicted by models simulating an entire NLR. Diagrams constructed from emission-line intensity ratios are widely used for this sort of comparison. Early empirical work by Baldwin, Phillips and Terlevich (1981) pointed out the importance of the [O III] λ5007/ Hβ vs. [N II] λ6584/ Hα diagram (hereafter the BPT diagram) for classifying emission-line galaxies as either active galactic nuclei (AGN) or star forming (SF). The optical diagnostic diagrams presented in Veilleux and Osterbrock (1987; hereafter VO87) added further important intensity ratios that are minimally affected by reddening. More recent work has extended this method into the ultraviolet (UV) and infrared (IR) (Spinoglio and Malkan 1992; Allen, Dopita & Tsvetanov 1998; Sturm et al. 2002; Groves, Dopita & Sutherland 2004a, 2004b (hereafter G04a, G04b)).

Figure 1 shows a BPT diagram with the sample of galaxies used here (see § 1.2) plotted as grey dots. We discuss the larger symbols and the labeled red lines below. The unlabeled blue lines show commonly used boundaries to separate different classes of objects. The blue dashed curve is the boundary between SF galaxies and AGN defined by Kauffmann et al. (2003), while the blue solid line is the theoretical upper limit on SF line ratios found by Kewley et al. (2001). An additional class of objects are the low-ionization galaxies called LINERs (Heckmann 1980). LINERs are characterized by very strong emission lines from neutral species, and are well separated from both AGN and SF galaxies in the BPT and similar line-ratio diagrams. The blue dotted diagonal line on Fig. 1 is the boundary between Seyferts and LINERS suggested by Kauffmann et al. (2003), although this definition was later refined (Kewley et al. 2006). Finally, there are other low-ionization galaxies called "transition" or "composite" objects which merge down into the SF galaxies on these line-ratio diagrams (Ho, Filippenko & Sargent 1993; Kauffmann et al 2003; Kewley et al 2006). These lie between the dashed and solid blue lines on Fig. 1. They may be a mix of SF, AGN and composite SF−AGN cases, introducing considerable confusion about the ionization mechanism over an extensive part of the full BPT diagram. Our study here includes an investigation of the relationship between these composite objects and typical higher-ionization Type II AGN.

We now turn to what is known about the physical nature of the NLR. There have been many studies using the observed intensity ratios to identify the ionization source in narrow line Seyferts and SF galaxies. Photoionization models have been remarkably successful in reproducing the observed emission-line ratios in high-ionization AGN (Ferguson et al. 1997 (hereafter F97); Komossa & Schulz 1997; G04b). The excitation mechanism for low-ionization narrow line regions in the region overlapping with SF galaxies remains uncertain and it is not even clear whether all such objects contain AGN. Photoionization models of AGN have focused on matching the emission from either high-excitation and low-excitation galaxies but none have attempted to simultaneously match both. Models that incorporate shock excitation (Dopita & Sutherland 1995, 1996) can account for LINERS but fail to fit high ionization AGN.

---

[1] Paper I designated emission lines by their vacuum wavelengths, but here we switch to the more commonly-used air wavelengths for transitions longward of 2000A.

The structure of the NLR is another open question. Three different models that can reproduce the observed emission from the NLR of Seyferts are: (1) a combination of matter and ionization bounded clouds (Binette et al. 1996); (2) radiation pressure dominated dusty clouds (G04a); and (3) locally optimally-emitting clouds (F97; see also Komossa & Schulz 1997).

The model that combines matter- and ionization-bounded clouds is parametrized by the ratio of the solid angles subtended by matter-bounded and ionization-bounded clouds. The matter-bounded component is responsible for high ionization lines, namely He II λ4686, and the ionization-bounded component is responsible for low to intermediate ionization lines. The dusty, radiation-pressure-dominated model is based on the principle that radiation pressure incident on grains is sufficient to moderate the density, excitation and surface brightness of high ionization regions. Finally, the locally optimally emitting cloud (LOC) model assumes that the emission lines come from a vast sea of clouds distributed over a wide range of densities and radial distances from a central source of ionizing radiation and that the observed emission-line intensity ratios are the result of a powerful selection effect: emission lines are produced in clouds that optimally emit them.

Each of these models can successfully represent the NLR in at least some galaxies, but it is not clear whether any of them can explain all NLRs. The low-ionization AGN, which we include in this present study, fall in a region of the BPT diagram that is also occupied by low ionization SF galaxies (cf. Tanaka 2012a, 2012b). This overlap with SF cases greatly confuses the issue of whether or not one general type of model can explain the full sequence from low-ionization non-SF objects to the much more luminous high-ionization objects such as classical Seyfert galaxies. The low-ionization AGN may or may not include cases where the ionization is due to a composite non-thermal and SF spectrum. What is needed is an unbiased set of AGN that span the full ionization range.

**1.2 The Active Galactic Nuclei sequence**

In Allen et al. (2013; hereafter Paper I) we developed a new technique that is able to separate AGN from SF galaxies with much greater purity than is usually achieved with older methods such as those based on BPT diagrams or principal component analysis. This MFICA technique allowed us to separate galaxies along a star-forming (SF) locus from those along an active galaxy (AGN) locus extending down to even quite low ionization levels. We used a sample of about $10^4$ low redshift ($0.1 < z < 0.12$) SDSS emission-line galaxies to isolate sequences, or "loci," of pure SF and of AGN cases. Full details of the MFICA analysis are given in Paper I; here we give a summary, focusing on the derivation of the AGN locus and of the increasing ionization sequence that lie along that locus.

All emission-line galaxies used in this study were selected from the redshift range $0.10 \leq z < 0.12$. Those used to generate the emission-line components, a sample of 727, were selected to have moderately high $r$-band signal-to-noise ratio ($16.0 \leq S/N < 23.0$) and positive equivalent width ($W_\lambda$) for each of H$\beta$, [O III] $\lambda$5007, H$\alpha$ and [N II] $\lambda$6584, with $S/N \geq 5.0$ for the flux measurements of H$\beta$ and [N II] $\lambda$6584. Their H$\alpha$ line widths were in the range $1.9\text{Å} \leq \sigma_{H\alpha} < 3.0\text{Å}$. The $\sim 10^4$ galaxies used to define the SF and AGN loci were allowed a wider range in $r$-band S/N ($15.0 \leq S/N_R < 30.0$) and line width ($1.9\text{Å} \leq \sigma_{H\alpha} < 3.5\text{Å}$), but otherwise followed the same criteria. This latter sample is plotted on Fig. 1.

Before analyzing the emission lines, an MFICA analysis was used to subtract the underlying stellar continuum from each galaxy spectrum. A sample of 393 galaxies, chosen to lie in the star-forming region of the BPT diagram as defined by Kauffmann et al. (2003), were then used to generate three emission-line MFICA components. These three components are sufficient to describe the emission-line spectra of star-forming galaxies, when combined linearly. The broader sample of 727 emission-line galaxies of all types was then used to generate a further two components; the combined set of five components allows the description of AGN spectra.

Figure 2 shows the spectra of these five emission-line MFICA components. The components were then fitted to the larger sample of $\sim 10^4$ galaxies. The effect of fitting the components is to project each galaxy onto the five-dimensional space of component weights, which we designate $W_1$–$W_5$. The weights were normalized such that they total unity for each galaxy. By performing this fit on a large number of galaxies we were able to examine their distribution in the space of component weights, which through the MFICA components themselves corresponds directly to their distribution of observed physical properties.

Galaxies that lie below the star-forming classification line of Kauffmann et al. (2003) are expected to be dominated by star formation; Stasińska et al. (2006) showed that no more than 3 per cent of their emission line flux arises from AGN. These galaxies form a tight "star-forming" sequence in the BPT plane (the "SF locus" shown on Fig. 1), and as expected this corresponds to a similarly tight sequence in the MFICA weights space.

Isolating a sample of pure AGN presented a greater challenge, because the BPT diagram does not allow for easy identification of a representative sample of AGN without significant contamination from star formation. However, the greater dimensionality of the MFICA weights does allow a sample to be identified. Figure 3 shows the two-dimensional projections involving the first three MFICA components (additional projections involving $W_4$ and $W_5$ are shown in figure 15 of Paper I). Fig 3 shows that the star-forming sequence (the blue line on the figure) has a significant contribution from component 2 for almost all of its length, excepting only the end at which it converges to $W_3 = 1.0$. Hence, galaxies with low values of $W_2$ can be expected to have emission lines with very little contribution from star formation. Indeed, there is a sequence of galaxies with low $W_2$ that also has strong contributions from components 4 and 5, indicating emission sources other than star formation. From inspection of the distribution of MFICA weights we chose $W_2 \leq 0.05$ as our first AGN criterion.

A second criterion is required in order to define the low-ionization limit of the AGN sequence

(defined below), *i.e.* how close to the star formation sequence it is allowed to extend. To this end, we required that $W_4+W_5 \geq 0.18$. The value of the cutoff was selected by trying a range of values, constructing an AGN locus for each one, and attempting to reconstruct the input sample in terms of summed contributions from the AGN and star formation loci. For cutoff values greater than 0.18, there was a population of galaxies that could not be described by a combination of the two loci, indicating that the AGN sequence was insufficiently long in those cases. When a cutoff value of 0.18 was used, these galaxies could be described by a combination of contributions from the two loci. Reducing the cutoff value below 0.18 produced no further improvement, so 0.18 was adopted for the following analysis.

Applying the above criteria leaves 5519 galaxies dominated by star formation, and 379 dominated by an AGN, with the remaining 4221 galaxies from our overall sample falling outside the BPT SF region but not classified as AGN because either W2 > 0.05 or (W4+W5) < 0.18. Of these rejected galaxies, 986 (23.3%) have an AGN contribution (ie W4+W5 >0.18) but W2 > 0.05.

The distribution in the five-dimensional space of MFICA weights was then parametrized separately for the SF and AGN-dominated subsamples, using the algorithm described in Newberg & Yanny (1997; hereafter NY97). The algorithm describes a distribution of points as a locus made of many segments. Each segment is a prism with an elliptical cross-section, or a higher-dimensional equivalent, surrounding a central "locus point". Starting from a pair of endpoints specified after visual inspection of the distribution, the algorithm adds segments one by one, stopping when the distance between the existing locus points is less than a fixed multiple, $N_{\sigma_{\text{spacing}}}$, of the width of the locus. We used $N_{\sigma_{\text{spacing}}} = 3.0$, the typical value used in NY97. We added an additional criterion that a new locus segment would only be inserted between two existing segments if there were at least 20 data points between the two existing locus points. This criterion improved stability in regions where the data points were sparser. The NY97 algorithm allows for a maximum distance between data points and locus points to be set, beyond which the data points do not contribute to the position of the locus point or to the size of the cross-section. For the distributions examined here, this criterion was found to make no significant difference to the parametrized locus, so it was not applied.

As well as defining the central spine of locus points, the NY97 algorithm defines a (possibly higher-dimensional) elliptical cross-section around each locus point, which describes the scatter of data points in that region. The algorithm defines the axes of the ellipse to be equal to the root mean square (RMS) of the distance between the spine and each data point. In this work, we multiply the RMS values by 1.5 to encompass a greater fraction of the data points within the locus, and refer to these increased values as the locus width.

With these techniques, we identified two statistically independent loci of observed galaxies, threading through the 5-dimensional component space. We called these the "SF" (star-forming) and "AGN" loci. Those loci are shown on Fig. 3. Combinations of spectra from the SF and AGN loci can reconstruct the spectra of galaxies that lie between the two loci. There is considerable separation between the loci, except at the very lowest-ionization end of the AGN locus. This degree of separation shows that AGN galaxies fall along a single locus suggesting that their variation can be represented by a single free parameter. We refer to the sequence of galaxies that lie at points along the AGN locus with increasing values of that free parameter as the "AGN sequence".

We measured emission line intensity ratios for subsamples of galaxies lying along the SF and AGN loci (§2, below) and mapped the loci back onto the BPT diagram in Fig. 1, where they are shown as the two red curves labeled "SF locus" and "AGN locus". The squares on Fig. 1 and Fig. 3 represent the positions of the individual galaxy subsets that lie on the AGN locus, while smaller triangles and diamonds show the positions of two parallel sequences lying at either end of the major axis of the ellipses, representing the error, as identified by the NY97 algorithm. We also show the position of a low-ionization subsample (called s01) on the SF locus, which we will use later in the paper (§5.4). On Fig. 1 and other line-ratio diagrams, the AGN locus does not pass through the realm of LINERs. Rather it starts out below the LINER region in the area occupied by transition objects, and then angles up into the AGN region. Fig. 1 clearly shows that the [O III]/H$\beta$ ratio is one parameter that locates galaxies along the AGN locus and therefore that the ionization parameter is a key factor in the definition of the AGN sequence.

### 1.3 The Goals of This Paper

The MFICA technique has picked out a distinct set of galaxies – the AGN sequence – which are different from purely SF galaxies, but includes only some of the objects in the AGN and composite regions of the BPT diagram. This sequence appears to have special significance in a statistical way, so we want to understand its physical interpretation. One end of the sequence contains high-ionization NLRs that are unambiguously classified as AGN, but the AGN sequence descends down into the low-ionization transition object region in a way that rejects SF galaxies with similar emission-line spectra. Since the AGN sequence is mainly a monotonic progression along a single AGN locus, we investigate the possibility that a single physical parameter, one that is more fundamental than just the resulting ionization level of the NLR, is responsible for the position of an object along the sequence. This defines the goal of this paper: we seek to find a single tunable physical parameter responsible for the variation in AGN ranging from Seyferts all the way down to the lowest-ionization AGN.

As a check on this, we will also initially include objects that represent the observed scatter orthogonal to the AGN sequence. However, the measured emission-line ratios described below show that there is little spread in properties perpendicular to the AGN sequence, over the range in parameter space that the Newberg & Yanni (1997) algorithm has identified as lying within the group of AGN with little contribution from star formation.

The open squares on Fig. 1 define a sequence leading up from the H II region into the AGN region of the BPT diagram. Differences in galaxies' locations along this direction are often ascribed to differences in the amount of mixing between photoionization by hot stars and photoionization by an AGN continuum. Given that the MFICA technique can tell the difference between SF and AGN objects at points well down into what is called the composite region, we explore here the alternative that *all* objects on the AGN locus have NLR's in which the photoionization is dominated by an AGN-like ionizing spectrum, and try to determine if some other underlying physical variable causes the range in properties along the AGN sequence.

In § 2 we describe composite observed spectra formed at five roughly equally spaced points along the MFICA AGN locus, and at two additional points to either side of the AGN sequence at each of those positions on the sequence. These composites range from very low ionization objects up through very highly ionized Type II AGN, and have sufficiently high S/N ratios that many additional observed line ratios involving weak lines can be measured as consistency checks in addition to using the usual strong line ratios which have been used to constrain previous models. None of these composite spectra show any evidence of a non-thermal continuum, except for their emission lines. Then in § 3, following F97 we adopt an LOC model and use it to reproduce the observational line-ratio diagrams. We investigate the sensitivity of the line ratios to various physical parameters. The physical interpretation of our results is discussed in § 4. This will include a comparison of the results from our LOC models to those for a purely empirical "mixing model" that combines spectra from SF galaxies and from a high-excitation AGN. Finally, § 5 summarizes our conclusions. We will address the SF locus in a future paper.

## 2. A comparison sample at representative points along the AGN locus

Composite spectra were formed from subsets of galaxies lying along the AGN sequence and to either side of it. They are named a$ij$, where the first index indicates the position along the AGN locus ranging from $i = 0$ at the low-ionization end to $i = 4$ at the high-ionization end. The second index is the position in a direction orthogonal to the AGN locus, with $j = 0$ corresponding to the "wing" closest to the SF sequence and $j = 2$ corresponding to "wing" further from the SF sequence. Fig. 1 shows where these points (shown as triangles) fall on the conventional BPT diagram. Note that on the BPT diagram the lowest ionization subsets (a0$j$) fall very close to the line representing the Kauffmann et al. (2003) lower boundary for finding AGN and are well below the line representing the Kewley et al. (2001) upper limit for SF galaxies. Table 1 lists some general properties of these 15 a$ij$ subsets, each of which contains a sample of the 50 galaxies whose MFICA weights most closely matched the selected points along the AGN locus. The $E(B-V)$ values were determined from the H$\alpha$/H$\beta$ intensity ratio as described below. We then list the observed emission-line luminosities $L$(H$\beta$) and $L$([O III] $\lambda$5007) and the continuum luminosity at $\lambda$5007, $L_\lambda$($\lambda$5007). These are followed by the corresponding dereddened values $L^c$(H$\beta$), $L^c$([O III] $\lambda$5007) and $L^c_\lambda$($\lambda$5007), and the observed equivalent widths W$_\lambda$ of H$\beta$ and [O III]. Finally, the table lists the relative weightings of each of the MFICA emission-line components when fitted to the coadded spectrum representing each individual subset along the AGN locus. Selected regions of the co-added spectra of the 5 central ($j = 1$) subsets are shown in Fig. 4. The additional subsets ($j = 0, 2$) are not shown in Fig. 4 due to their close similarity to the central subset.

We then measured emission-line strengths from the co-added spectrum for each subset. For most lines, we just integrated the flux within the line profile above a locally fitted continuum. The accuracy was about +/-5 per cent for the stronger lines ranging to +/-20 per cent for the weakest lines based on the S/N ratio in the adjacent continuum. We separated the [S II] doublet by simply dividing the blended profile at the lowest point between the two lines. He I 5876 is on the wing of Na D, which is the one absorption feature in the underlying galaxy spectrum that obviously did not subtract off properly along with the rest of the continuum. Our He I $\lambda$5876 measurements are based on fitting the He I line with the profile of the H$\beta$ emission line, with a typical uncertainty of about 10 per cent. Table 2a lists the observed (reddened) value, while Table 2b lists the dereddened values assuming for simplicity, a standard Galactic $R_V = 3.1$ reddening curve (Cardelli, Clayton & Mathis 1989) with $E(B-V)$ chosen to produce dereddened $I$(H$\alpha$)/$I$(H$\beta$) = 2.86, appropriate for Case B recombination at $n_e=10^2$ cm$^{-3}$ and $T_e=10^4$ K (Osterbrock & Ferland 2006; hereafter AGN3). The dereddened values are the ones used throughout the remainder of the paper. The S/N obtained from co-adding spectra allows weak lines such as [S II] λ4070 (the sum of the [S II] λλ4068,4076 doublet), [Ar IV] λ4711, [N I] λ5200, [N II] λ5755, He I λ6678 and [O II] λ7325 to be measured in most cases, providing important consistency checks on our models.

## 3. Dust-Free LOC Models

### 3.1 Method

We used the LOC model, which treats the NLR as the sum of a large number of separate gas clouds spread out around a central ionizing source (Baldwin et al 1995). Once the incident spectral energy distribution (SED) and the chemical abundances are set, the major free parameters in the LOC model are the distributions of the individual clouds in their radial distances $r$ from the source of ionizing radiation, and in their gas densities $n_H$.

The individual clouds were modeled using version 10.0 of the plasma simulation code Cloudy (Ferland et al. 1998). For simplicity, we assumed that there is no ISM attenuation of the AGN continuum radiation incident on the clouds. The consequences of this are described in F97. We used constant density models; the results of Pellegrini et al. (2007) show that constant-density and constant-pressure models give very nearly the same result for optical lines, which form in the warm gas within the $H^+$ zone. The distinction is mainly important for infrared lines which form in cool atomic gas and which are not considered here. For solar abundances, we used those summarized in G04a, taken from a series of papers by Asplund and his collaborators (Asplund et al. 2000; Asplund 2000, 2003; Allende Prieto et al. 2001, 2002). Abundances for additional elements not given by G04a are taken from F97. Following F97, we used an ionizing luminosity $L_{ion} = 10^{43.5}$ erg s$^{-1}$, typical of Seyfert galaxies. Our simulations proceeded until the electron temperature fell below 4000 K or above $10^5$ K. Gas below 4000 K does not significantly contribute to optical emission lines, although it does produce infrared lines, and gas above $10^5$ K contributes principally to X-Rays. These models did not include any cosmic rays. The Cloudy models all terminate before reaching molecular regions where cosmic ray ionization from the Galactic background becomes important. Komossa & Schulz (1997) studied the effect of adding the standard Galactic cosmic ray background to similar models, and found that the emission line ratios changed by only 1 per cent.

A series of grids of individual clouds were computed over a range of densities and radii that represent plausible values for gas in the NLR. Each grid consisted of 7171 cloud models. The radial distance from the ionizing source to the individual cloud, $r$, was varied in 0.1 dex steps from $10^{16}$ to $10^{23}$ cm and the hydrogen density, $n_H$, was varied in 0.1 dex steps from $10^0$ to $10^{10}$ cm$^{-3}$. We assumed that the ionizing flux was isotropic so that a difference in radii reflects a difference in flux proportional to $r^{-2}$. Differences in $L_{ion}$ would appear as a rescaling of the radial distances by $(L_{ion}/10^{43.5})^{1/2}$. The total hydrogen density was kept constant for each individual cloud but the molecular and electron densities were determined self-consistently and so vary with depth. The results from these grids are shown below in terms of the equivalent width of various emission lines, which indicates the efficiency of each cloud in reprocessing the continuum into a particular emission line. These equivalent widths are relative to the continuum level at 4860Å in the models, which includes only the ionizing AGN continuum, a quantity that is not directly measurable in the observations.

We computed both dust-free and dusty grids of clouds. By "dusty", we mean LOC grids that include dust in the individual clouds in parts of the NLR where sublimation is not expected to have occurred, while "dust free" refers to LOC grids with no dust anywhere. This current section describes the dust-free models.

The LOC model is based on the fact that different emission lines are optimally emitted in different individual clouds according to the correct density and incident flux for efficient production of each particular line, while we observe only the integrated spectrum of the ensemble of clouds. This gives rise to a powerful selection effect in which the overall radial and density distributions of the clouds largely determines the measured spectrum. As in F97 (see also Baldwin et al. 1995), we modeled the total emitted spectrum integrated over radial distance and density distributions defined as $f(r) \propto r^\gamma$ and $g(n) \propto n_H^\beta$, respectively, where $\gamma$ and $\beta$ are free parameters. An average spectrum over many galaxies as we have produced here is more likely to be correctly described by a broad power-law distribution than might be the case for an individual galaxy. The total luminosity of the line is then,

$$L_{\text{line}} \propto \iint r^2 F(r, n_H) r^\gamma n_H^\beta \, dn_H \, dr \qquad (1)$$

where $F(r, n_H)$ is the flux of the line and $r^\gamma n_H^\beta$ is the spatial distribution function (Bottorff et al. 2002). Throughout the remainder of the paper, when the log of densities and radii are given, the units are $cm^{-3}$ and cm, respectively. We chose the integration limits to include the range of parameter space that is physically relevant for the NLR. Gas with densities $\log(n_H) > 8$ is above the maximum critical density for optical forbidden lines. Gas at radii $\log(r) < 17.48$ would either be much too hot to produce strong optical emission lines, or would have high density which would suppress the forbidden lines. Gas with densities $\log(n_H) < 2.0$ does not optimally emit many lines, the notable exception being [S II] λ6720. Finally, gas with $2\log(r)+\log(n) < 41.5$, which roughly corresponds to $\log(U) > 0.4$, was not included because it is so highly ionized that the gas that produces the observed emission lines is essentially transparent to ionizing radiation. This is the justification for the integration limits used by F97.

Here we present LOC integrations within the limits $2.0 < \log(n_H) < 8.0$ and $17.48 < \log(r) < 22.0$. Although our grids of models extend out to $10^{23}$ cm, the majority of the equivalent widths considered in this work peak at much smaller radii. Our adopted limit of $10^{22}$ cm is the same one used by F97. In addition, the 3" diameter SDSS fiber translates to approximately $10^{22}$ cm for galaxies observed at $z \sim 0.1$. We experimented with changing the integration limits to include a larger range, but this only produces small changes to the integrated spectrum. Changes in the abundances, the SED, $\gamma$ or $\beta$ produced more noticeable effects.

We compared the observed integrated predictions to the predictions by using line-ratio diagrams as in Baldwin, Phillips & Terlevich (1981), Veilleux & Osterbrock (1987), and elsewhere. We discuss these diagrams below. We will exclusively describe dust-free LOC grids until §3.5 where we address the fact that our dusty models do not fit some of the AGN subsets as well as our dust free models.

## 3.2 SED Optimization

The intrinsic ionizing SED in AGN is uncertain. The various forms of the incident continuum assumed in previous work include simple power laws, $f_\nu \propto \nu^\alpha$, (Binette et al. 1996, G04a), a combination of power law and blackbody continua (Komossa & Schulz 1997) and multi-component distributions (F97). In addition, the SED viewed by the BLR (Korista et al. 1997) may be different than the SED viewed by the NLR, and parts of the SED, such as emission originating from a dusty torus are only visible for AGN at certain viewing angles. We adopted a multi-component approach as in F97 and used He I/H I line intensity ratios to infer the SED.

We considered the three different spectral energy distributions (SEDs) shown in Fig. 5. They all have the general form

$$f_\nu \propto \nu^{\alpha_{\rm uv}} e^{-h\nu/kT_{\rm cut} - kT_{\rm IR}/h\nu} + C\nu^{\alpha_x} \qquad (2)$$

where $\alpha_{\rm uv}$ is the low-energy slope of the "big bump," $T_{\rm cut}$ is the UV temperature cutoff, $T_{\rm IR}$ is the IR temperature cutoff, $C$ is a normalizing constant, and $\alpha_x$ is the slope of the X-ray component. The factor $C$ sets the scaling between the big bump and the X-ray power law, and in Cloudy is determined by specifying $\alpha_{\rm ox}$, which is given by

$$\frac{f_\nu(2\ {\rm keV})}{f_\nu(2500\ \text{Å})} = 403.3^{\alpha_{ox}} \qquad (3)$$

(Korista et al. 1997). The X-ray component (last term of equation 2) was assumed to fall off as $\nu^{-2}$ above 100 keV.

A commonly used AGN SED is given by Mathews & Ferland (1987; hereafter MF87), and is shown as a dotted line in Fig. 5. A combination of direct observations, with inferences based on the emission-line spectrum, determined the components of this SED. The X-ray slope of the SED, $\alpha_x \sim -0.7$, is typical for radio-loud AGN (Zamorani et al. 1981). As we describe below, this has a significant effect on the equivalent width contours but minimally affects the LOC integrations. F97 used a modified SED, shown on Fig. 5 as the dashed line. This SED differs from that of MF87 in that it lacks infrared emission from a dusty torus, includes a soft X-ray component, and has a softer X-ray slope $\alpha_x \sim -1.0$ that is typical of radio quiet quasars and AGN (Elvis et al. 1994).

The He II λ4686/ Hβ intensity ratio is an important test of these possible SEDs, because (1) He II and Hβ are recombination lines and therefore less sensitive to $T_e$ than collisionally excited lines (AGN3); (2) the ratio is relatively insensitive to large changes in the ionization parameter (Binette, Courvoisier & Robinson 1988). Photoionization models of AGN often underpredict this ratio (Ferland & Osterbrock 1986).

Using the F97 and MF87 SEDs, we computed two separate grids of Cloudy models as outlined in §3.1, employing a solar composition for this preliminary step, and then ran LOC integrations. The predicted [O III] λ5007/ Hβ vs He II λ4686/ Hβ line ratio diagrams for our dust-free LOC models using each of these SEDs are shown in the upper two panels of Fig. 6 as colored lines connecting a series of circles. Each line represents the LOC results for a different density weighting in the range $-1.8 \leq \beta \leq -0.6$ in increments of 0.4 with $\beta = -1.8$ [blue], $\beta = -1.4$ [green], $\beta = -1.0$ [red] and $\beta = -0.6$ [cyan]. Along each individual line the radial weighting is varied over the range $-2.0 \leq \gamma \leq 2.0$, and the circles along the line represent different LOC models run at increments in $\gamma$ of 0.25. The largest solid circles indicate the most negative radial distribution of clouds, $\gamma = -2.0$. The circles with white centers represent models using our best-fitting set of the remaining free parameters, as described below in §3.4. The black diamonds, triangles and squares represent the ratios measured from the AGN observations, with each sequence in the observations indicated by a different line style and symbol. The largest of these symbols represents the top of the AGN locus for that sequence. The largest square should be compared to the largest white-filled circle, which represents our best-matching set of free parameters for the high ionization end of the observed sequence.

Fig. 6 shows all three observed AGN sequences (the central a$i$1 locus and the two "wings" corresponding to a$i$0 and a$i$2). The three observed sequences are very similar on Fig. 6. In fact, even in the full set of line ratio diagrams described below the difference between each sequence is so slight that we cannot determine a physically meaningful interpretation for the wings; they appear to just represent object-to-object scatter about the central locus. For this reason, we compare our models only to the central locus and on subsequent line-ratio diagrams show only the observed points for the central AGN locus.

We explored the entire range of $\beta$, $\gamma$ and SED presented in Fig. 6 when fitting the subsets within the central locus. Our goal, as presented in the introduction, is to find a single tunable free parameter responsible for the systematic variation of AGN along the BPT diagram. As is discussed below (§3.4), varying only the radial distribution parameter $\gamma$ of clouds while holding the density weighting parameter β constant provides a reasonable fit to the AGN locus. This became apparent at an early stage in the investigation and so we used this approach to optimize the SED.

It can be seen that the higher-ionization observed points in the upper two panels of Fig. 6 are in good agreement with the calculated curves. To determine the best set of free parameters, we varied the radial distribution of clouds, starting with the top of the locus, trying to reproduce the highest ionization subset to within a factor of 2, before proceeding to the next subset. Here and throughout the paper, we adopt a factor of 2 as our criterion for an acceptable fit between the predicted and observed line ratios. This is principally due to uncertainties in the models after taking into account the uncertainty in the SED and abundances, and the fact that the LOC models just use simplified power laws to describe the distributions of $r$ and $n_H$. As mentioned in §2, the measurement of an emission line ratio containing two weak emission lines would possess an uncertainty of ~30%. Thus, the observational uncertainty is less than the systematic uncertainty in our models.

For the F97 SED (Panel (a) of Fig. 6), the models producing integrated spectra with moderate [O III] λ5007/ Hβ ratios (moderate ionization) overpredict the He II/ Hβ ratios by almost a factor of three. In contrast, the MF87 SED produces a He II/ Hβ ratio within a factor of 2. This is closely related to the "Stoy method" of determining the SED (AGN3) and we adjust its hardness to reproduce the ratio.

We optimized the SED to produce a third continuum shape, shown on Fig. 5 as the solid line labeled "optimized, dust-free", to try to produce a better fit to the measurements from our galaxy sample. In addition to a SED similar to radio-loud quasars (MF87), we wanted one that was typical of AGN (F97) but that still matched the moderate ionization He II/ Hβ ratio. To achieve this, we modified the F97 SED by shifting the "big bump" until the He II/ Hβ ratio matched the moderate to high ionization subsets. This modified "bump" turned out to match the one in the MF97 SED. We also kept the $\alpha_x$ to provide closer agreement with the mean value of -1.0 found in radio quiet quasars and AGN (Zamorani et al. 1981; Steffen et al. 2006); this makes the optimized SED significantly different from the MF87 SED in the x-ray region.

Our optimized SED is characterized by $T_{cut} \sim 4.2 \times 10^5$ K, $\alpha_{UV}$ = -0.3, $\alpha_{ox}$ = -1.35, $\alpha_x$ = -1.0, and $kT_{IR}$ = 0.14 eV. The bottom panel of Fig. 6 shows the resulting predicted line-ratio diagrams and verifies the fit to the moderate and high-ionization He II/ Hβ ratio. Further adjustments were made to match the lower ionization He II/ Hβ ratios but we were unable to simultaneously account for both the lower and upper end of the locus.

Fig. 7 shows the $W_\lambda$ distributions for several emission lines for each of the three SEDs discussed so far. We note that the peak emissivity of some strong lines occurs at two locations on the LOC plane for the Mathews & Ferland (1987) AGN continuum, but the Ferguson et al. (1997) continuum only creates a single maximum. We found that this is caused by the different assumptions in the form of the X-ray continuum. These second optimally emitting regions minimally affected the LOC integrations. Furthermore, the densities at which the second optimally emitting regions occur are most likely larger than those physically relevant to the NLR. We mention this result here because X-ray effects are important in other contexts of AGN (e.g. in the BLR).

The MF87 and our optimized SED give essentially identical fits to the observed points on Fig. 6, which are slightly better than the fit for the F97 SED. We chose to use the optimized SED for the remainder of our analysis.

### 3.3 Metallicity Effects

We next adjusted the metallicity. We found that [N II] λ6584 was underpredicted with our solar abundance set, so we adjusted the nitrogen abundance. Nitrogen abundances have long been known to deviate from solar abundances in AGN (e.g. Storchi-Bergmann & Pastoriza 1990; Dopita & Sutherland 1995; Hamann & Ferland 1999). Baldwin et al. (2003) found that nitrogen abundances a factor of 15 above solar are needed to accurately represent the QSO Q0353-383, and Bradley, Kaiser & Baan (2004) found a factor of 3–4 enhancement in the weak NLR in M51. Nitrogen enhancement in AGN is thought to stem from secondary CNO nucleosynthesis, where carbon and oxygen are pre-existing, and then distributed by stellar winds from massive stars. This creates a unique scaling relationship with metallicity, [N/H] $\propto Z^2$ while other elements scale linearly with metallicity (Hamann & Ferland 1999).

We used that scaling relation to tune the abundances to fit [N II] λ6584/ [O II] λ3727, a metallicity indicator (Groves, Heckman & Kauffmann 2006) and [N II] λ6584/ Hα. After iterating over a range of metallicities[2], we found that a metallicity of 1.4 Z$_\odot$ provides the best agreement with these two line ratios. Enhancing only nitrogen by a factor of 1.5 above solar would give an equally good fit to the observations, but we used the 1.4 Z$_\odot$ value appropriate for secondary enrichment.

---

[2] Here we adopted the common usage of the term "metallicity" as the metals abundance relative to hydrogen, by number, normalized to solar values. However, we note that metallicity is more properly defined as a mass fraction.

This abundance set enabled all but the lowest AGN subset to be fitted, indicating that a single metallicity is sufficient to explain AGN over at least most of the AGN sequence. We did not assume this fact prior to any modeling, but this result is physically consistent with the observations given that the galaxies in the sample span a narrow redshift and apparent magnitude range, and hence on average are likely to have similar chemical enrichment.

Our final abundance set for the dust-free models, by number relative to hydrogen, is given in the log as follows: He: -0.987; Li: -8.54; Be: -10.4; B: -8.97; C: -3.46; N: -3.90; O: -3.16; F: -7.37; Ne: -3.77; Na: -5.53; Mg: -4.27; Al: -5.36; Si: -4.34; P: -6.28; S: -4.65; Cl: -6.57; Ar: -5.45; K: -6.72; Ca: -5.50; Sc: -8.65; Ti: -6.81; V: -7.83; Cr: -6.16; Mn: -6.32; Fe: -4.39; Co: -6.93; Ni: -5.60; Cu: -7.58; Zn: -7.19.

### 3.4 Comparison to Observed Diagnostic Diagrams

Our goal in this paper is find a single tunable physical parameter responsible for the variations along the AGN sequence ranging from Seyferts all the way down to the lowest-ionization objects. In this subsection we compare observed line ratios to those predicted by the dust-free LOC models described above, and show that with our adopted SED and abundances we can achieve reasonable agreement along the AGN sequence by varying only the LOC radial distribution parameter $\gamma$.

In order to get a fuller picture of the physical conditions in the individual clouds as a function of their position on the $\log(n_H)$–$\log(r)$ plane, Fig. 8 shows the ionization parameter, $U$ (dotted lines), and the gas temperature in the [O III]-emitting zone $T_e$ (dashed and solid lines) at the illuminated face of the cloud. Note, while models are shown that are either above $10^5$ K or below 4000 K, these models were terminated after one zone and therefore have negligible contributions to the overall spectrum. The cutoff for detectable lines is taken to be EW = 1Å, which roughly corresponds to $\log(U) = 0.4$. Each $U$ contour has a slope of $(n_H dr)/(r dn_H) = -0.5$. Fig. 9 displays the equivalent width contours for 20 important emission lines, as a function of $\log(n_H)$ and $\log(r)$.

Next, Figs. 10-14 show 17 different diagnostic diagrams (but with one diagram repeated twice to make 18 total) predicted for the LOC integrations using the dust-free model, and compare them to the observations in the same way as was done in Fig. 6. The lines and symbols in Figs. 10-14 are keyed the same as in Fig. 6, with the black lines and symbols showing the observed line ratios for the central AGN locus, and the colored lines and symbols showing predicted line ratios, with a separate predicted line for each sequence of LOC models in which $\beta$ is held constant but $\gamma$ is varied. Each panel is on the same scale to emphasize at a glance which line ratios are better indicators. The error bars in the bottom right corner of the last panel in each figure represent a factor of two error. These diagrams all place useful constraints on the values of $\beta$ and $\gamma$ used to fit the individual observed subsets. For completeness, the appendix shows 7 additional diagrams that do not constrain the models. We show this large number of diagnostic diagrams in order to give the full picture of how well our model actually fits *all* of the data, rather than just a few of the typical, more easily-measured line ratios.

We discuss the individual diagrams from Fig. 10-14 below, but a first glance over the full set of diagrams shows that our LOC models can fit almost all line ratios for the upper half of the observed AGN locus (the black points starting from the larger black symbols and working back down the observed sequence). The larger black square represents the observed extreme AGN case, a41.

Examination of Figs. 10–14 shows that the observed line ratios tend to pick out an overall sequence in γ (the radial dependence parameter), but not in β (the density parameter). The lower ionization subsets of emission line ratios such as, [Ar IV] 4711 / [Ar III] 7135, [O I] λ6300 / Hα, and [O I] λ6300 / [N II] λ6584 point to more negative density weightings, while [S II] λ6720 / Hα, He II λ4686 / Hβ, and [O III] λ4363 / λ5007 point to the opposite.

We used the same method as outlined in §3.2 to determine the set of radial distributions that best fit our observed subsets. As before, we started at the high ionization portion of the locus and found the γ value that fit the greatest number of emission line ratios to within a factor two. More emphasis was placed on satisfactorily fitting strong line ratios, and those more established in the literature, as opposed to ratios drawing upon weaker lines, which are often degenerate in our parameter space.

The LOC model that provides the overall best fit to the observations of the highest-ionization AGN subset has β = -1.4, γ = -0.75 and is indicated by the larger open (filled in white) green circle. The observed a41-a01 sequence is matched using a density weighting of *β* = -1.4 (the green line on Figs. 8-12) while changing the radial weighting as parametrized by *γ* in Eq. 2. The a41-a31-a21-a11-a01 observations are fitted to reasonable accuracy in most panels on Fig. 10-14 by the LOC models with values *γ* = -0.75, *γ* = -0.5, *γ* = -0.25, *γ* = 0.0, *γ* = 1.0 respectively, each of which is shown as an open circle. A perfect fit would have the larger black square fall exactly on top of the larger open circle, and then to have each successive black square fall on top of each successive open circle.

Table 3 shows the line ratio predictions for our best-fitting model for each observed subset as well as a comparison to the observed value. At high - moderate ionization, the fits are usually within a factor of two. This demonstrates the effectiveness of the MFICA method at picking out an AGN ionization sequence that is largely described by the variation of a single physically meaningful parameter – the radial distribution of clouds.

The line ratio diagrams shown in Figs. 8–12 are organized according to which physical parameter they do the best job of constraining. Next, we discuss the details of the comparisons between the models and the observations in terms of these physical parameters.

*3.4.1 Excitation Mechanism*

Panels (a)-(d) of Fig. 10 show classic diagnostic diagrams from BPT and VO87, which are particularly sensitive to whether the excitation source is a power law (AGN) or starburst. The degree to which our model fits the AGN observations emphasizes its consistency with line ratios widely used to identify the excitation source and classification of emission line galaxies. Comparing simulations with power law spectra and starlight to observations resulted in the development of these diagrams.

We used these diagrams to make a baseline assessment of our model, since we must be able to reproduce typical emission line ratios before exploring other diagnostic diagrams. Using the criterion of a factor of two representing an adequate fit, we successfully matched [N II] λ6584/ Hα over the entire AGN sequence, [O III] λ5007/ Hβ, [O II] λ3727/ [O III] λ5007 and [O I] λ6300/ Hα for the a41-a31-a21-a11 sequence (higher to low ionization subsets), and [S II] λ6720

/ Hα for the a41-a31-a21 sequence (higher to moderate ionization subsets). These results show that we have sufficient agreement with established excitation diagnostics to proceed on to other line ratio diagrams.

### 3.4.2 Spectral Energy Distribution

Although we have already discussed the SED optimization in terms of fitting the [O III] λ5007/Hβ vs He II λ 4686/Hβ diagram we display this diagram again in Fig. 11(a) but this time for our model that has the increased metallicity described in §3.3. It is also important to demonstrate that our final model fits other line ratios that are also sensitive to the SED; Fig. 11 (b)-(c) show that this is in fact the case. This adds important cross-checks to the results from the usual strong-line diagrams in previous investigations. In order to limit the effects of differing abundances, these diagrams mostly use ratios of lines from different ionization states of the same element. The S/N obtained from co-adding spectra instead of using emission lines from individual AGN makes these diagrams more trustworthy indicators.

Fig 11(b) combines a number of emission lines from different ionization states but for the same element. Our model matches both of these line ratios to within a factor of two for the entire length of the locus except for the lowest ionization subset of [O II] λ3727 / [O III] λ5007. Panel (c) again shows [O II] λ3727 / [O III] λ5007 but now coupled with He II λ4686 / He I λ5876, in which the emission lines come from ions with very different ionization potentials. This emission line ratio agrees with observations along the entire length of the locus.

### 3.4.3 Temperature and Density

Fig. 12 probes the average temperature and density. Except as noted, these figures use standard temperature- and density-sensitive intensity ratios (see AGN3, chapter 5). In the case of the temperature sensitive [O III] λ4363/ [O III] λ5007 ratio (Panels (a)-(b)), the models come close to matching the higher ionization subsamples but the lower ionization subsamples give higher temperatures than those given by our models. Panel (c) confirms this finding using another temperature indicator, [N II] λ5755/ λ6584. The observations agree with our models at the higher ionization end of the sequence but at lower ionization the ratio disagrees with our models by indicating higher temperatures than predicted. This is an important discrepancy that we will explore further in §4.2.

Panel (a) shows a typical density sensitive ratio, [S II] λ6716/ λ6731, which is in excellent agreement with our models. This ratio is very insensitive to our free parameters, which is not surprising in these LOC models because the density indicated by the [S II] line ratio is close to the density for optimal emission of these lines. Panel (d) shows a diagram originally proposed by BPT, and therefore also appears in Fig. 10, but which is a good indicator of the ionization parameter (Komossa & Schulz 1997; G04b). The observed [O II] λ3727/ [O III] λ5007 agrees nicely with all but the lower end of the sequence and even in this region only disagrees by a factor of 3-4.

### 3.4.4 Dust Indicators

The diagnostics in Fig. 13 are taken from G04b, who found these line ratios to be sensitive to the radiation pressure from dust grains and to emphasize the success of their dusty model. However, we find that our dust-free models match the observations just as well as the G04b dusty models. For example, Panel (a) shows [N I] λ5200 and [O II] λ7325, which are typically weak in Seyfert galaxies, and there is decent agreement between the AGN observations and our dust free model. G04b found dramatic differences between the observations and dust-free models on this diagram. G04a argued that the diagram in Panel (b) ([O II] λ3727/ [O III] λ5007 vs. He II λ4686 / Hβ) points out the success of their dusty models over their dust-free models at high ionization parameters. In contrast, our dust-free models fit quite nicely over a large range of ionization. The integrations in Panel (c) match our high to moderate ionization region of the sequence. Panel (d) was used by G04b to emphasize that their dusty model is not devoid of problems because of the inability to reproduce high temperatures at moderate ionization; our model shows that this problem is also present in our dust-free models.

It should be noted that the classical Seyferts to which G04b compared their models almost all lie above the dotted and solid lines in Fig. 3, so that in terms of the BPT and VO87 diagrams their results should be compared only to our 3 highest-ionization levels. These are the cases for which our dust-free models give the best fits to the observations.

*3.4.5 Abundances*

Fig. 14 shows ratios of lines from ions with similar ionization potentials, but from different elements, leading to abundance sensitivity. He I λ5876 depends linearly on the helium abundance; therefore Panel (a) shows that our default solar helium abundance is correct for all but the lower ionization parts of the sequence.

Panel (b) displays the main determinant in our metallicity optimization, [N II] λ6584/ [O II] λ3727, and our model fits the entire sequence to within 10 per cent except for the lowest ionization observation. The diagram in Panel (c), [Ne III] λ3869/ He II λ4686 vs. [O III] λ5007/ [Ar III] λ7135 uses lines with fairly high ionization potentials, and shows agreement between the models and observations over almost the entire length of the AGN locus. While it was possible that the chemical abundances might vary along the AGN sequence, the results here indicate that a constant metallicity is a reasonable approximation.

## 3.5 Predicted $L_{ion}$ and NLR sizes

The combination of the observed Hβ luminosities and the LOC models place limits on the ionizing luminosity and the NLR size. The predicted emission-line luminosity for a LOC model with a total covering factor Ω is,

$$L(\text{line}) = \Omega L_\lambda \frac{\sum W_\lambda r^\gamma n_H^\beta}{\sum r^\gamma n_H^\beta} \quad (4)$$

where $L_\lambda = dL(\lambda) / d\lambda$ is the AGN continuum luminosity at 4861Å, $W_\lambda$, $r$ and $n_H$ are the equivalent width, radius and density. The sums are over the integration limits presented in §3.1.

The total equivalent widths, $W_\lambda(line)$, predicted from our model are then simply,

$$W_\lambda(line)/\Omega = \frac{\sum W_\lambda r^\gamma n_H^\beta}{\sum r^\gamma n_H^\beta} \quad (5)$$

and are on the order of $W_\lambda(H\beta)/\Omega \sim 200$ Å, relative to the unobscured AGN continuum. These are listed in Table 3. In principle, we could compare the observed $W_\lambda(H\beta)$ to this value to determine the covering factor. The observed $W_\lambda(H\beta)$ relative to the total observed continuum is approximately the same for all subsets, with $W_\lambda(H\beta) \sim 5$ Å (Table 1). However, the observed continuum is completely dominated by starlight, and we cannot detect the AGN contribution down to an estimated upper limit of 1% of the total continuum level. Accounting for this would scale up the observed $W_\lambda(H\beta)$ by as much as a factor 100. On the other hand, the AGN continuum is likely obscured in the Type 2 AGN, which would work in the opposite direction in terms of trying to estimate the covering factor. In conclusion, we do not have useful observational limits on the covering factor.

For each subset, $L(H\beta)$ is measured (Table 1) and the best-fitting LOC models give the values under the sums in Eq. (4). This means that we can solve Eq. (4) for $\Omega L_\lambda$ and hence $\Omega L_{ion}$, and also for the scaled maximum radius included in the LOC model, $r_{max} = 10^{22} (L_{ion}/10^{43.5})^{1/2}$ cm.

Changing the radial concentration of clouds also changes the effective size of the NLR. We define a characteristic radius for a given emission line by

$$r_{\text{char}} = \frac{\sum_{i=0}^{N} r W_\lambda r^\gamma n_H^\beta}{\sum_{i=0}^{N} W_\lambda r^\gamma n_H^\beta} \quad . \qquad (6)$$

Table 3 lists values of $L_{ion}\Omega$, $r_{max}\Omega^{1/2}$, $r_{char}(H\beta)\Omega^{1/2}$ and $r_{char}([O\ III]\ 5007)\ \Omega^{1/2}$ for each model. The deduced range in $L_{ion}$ and also in the NLR sizes $r_{char}$ depend on the covering factor, which cannot be measured. For a covering factor $\Omega = 0.4$, which is typical for the obscuring torus (Rowan-Robinson et al. 2008), the values of $L_{ion}$ range from $3$–$16 \times 10^{43}$ erg s$^{-1}$. These are typical for luminous Type 2 AGN. Fig. 15 plots $r_{char}$ (now shown in units of kpc) against the radial concentration, for [O III] $\lambda 5007$ and H$\beta$ with $\Omega = 0.4$.

## 4. Dusty Models

In addition to the dust-free models we also ran grids of dusty models, to test whether or not a dusty NLR could do an equally good job of reproducing the observations along the entire length of the AGN locus, again by tuning only a single free parameter. A dusty NLR model was also explored by F97 and preferred by G04a, making it natural to see if such a model is a viable alternative to a dust-free interpretation.

Dust is difficult to accurately model in a physically consistent way, since sublimation occurs at small distances from the central object, so following F97 we adopted a step function in our dusty grids to artificially account for this process. We used a mix of graphite and carbonaceous grains and a grain size distribution typical of an H II region (Baldwin et al. 1991). At $r \leq 10^{16.9}$ cm the grain temperature is high enough to sublimate both types of grains so our models were dust-free with solar abundances. For $10^{16.9} < r \leq 10^{17.6}$ cm, the silicates are too hot to exist, so we included only graphite grains and used a solar composition, except for carbon which was depleted by 15 per cent. Finally, for $r \geq 10^{17.6}$ cm, we include both graphite and carbonaceous grains and adopt Orion Nebula abundances (Baldwin et al. 1996) that are typical of an H II region. This is the procedure used by F97. In reality, different sized grains sublimate at different temperatures. However, we followed F97 and ignored this effect, assuming that there is threshold radius for each grain composition after which those particular grains are allowed to form.

As with our dust-free models, the He II λ4687/ Hβ ratio was overpredicted using the F97 SED. We optimized the SED using the same process as discussed in §3.2 and arrived at an incident continuum characterized by $T_{cut}$ ~ $1.5 \times 10^5$ K, $\alpha_{UV}$ = -0.5, $\alpha_{ox}$ = -1.4, $\alpha_x$ = -1.0, and $kT_{IR}$ = 0.14 eV (shown on Fig. 5 as the dash-dotted line).

Depending on the radius, our baseline abundance set changes due to the formation of grains, therefore there are actually slightly different metallicities within a single dusty grid. As with the dust-free models, there was the need for an increase in metallicity or selective nitrogen enhancement. We scaled the abundances using the empirically determined scaling relationships given in §3.3 and found that a metallicity of 1.4 $Z_\odot$ in the dust-free portion of the grid gave a satisfactory fit.

We then carried out the same analysis as for the dust-free models. Table 4 provides the emission-line ratio predictions and their comparison to our subsets for our best dusty model, in the same format as Table 3. Our dusty model fits a few line ratios slightly better in the moderate ionization part of the sequence. In particular, the entire [O II] λ7325/ Hα sequence and the moderate ionization [O III] λ4363 / [O III] λ5007 subsets are fit within a factor of two.

However, our dust-free model fits all but the moderate ionization subsets of [O II] λ7325/ Hα to within a factor of three. Furthermore, our dusty models still have the same problem as the dust-free models in that they predict that the [O III] λ4363 / [O III] λ5007 line ratio should get larger as the ionization level increases, while the opposite is observed to happen. The dusty models predict a systematically higher [O III] λ4363 / [O III] λ5007 emission line ratio than do the dust-free models, as a result of the gas being hotter due to photoelectric heating by dust (see Baldwin et al 1991), so the subsets are matched at a lower observed ionization level. Our dusty model similarly fails to reproduce the observed [N II] temperature at high ionization.

Our dusty models do not reproduce the lower ionization portion of some of the VO87 diagrams as well as our dust-free models and fail to account for the high ionization observations in a few ratios. Fig. 16 shows some line ratios, which are not fit as well in the dusty case as in our dust-free model. Fig. 16a should be compared to Fig. 10b and Fig.16b to Fig. 13a. The dusty case is not ruled out by these results, but the dust-free models give a marginally better fit at higher ionization.

## 5. Discussion

### 5.1 Physical Meaning of the AGN Locus

Analysis of the AGN region of the BPT diagram has historically been limited to well-identified emission-line galaxies displaying strong AGN properties. F97 studied these galaxies using LOC models, and Komassa & Schulz (1997) also studied luminous, high ionization Seyfert 2 galaxies using similar composite models having clouds spread over wide ranges in radial distance and density. In this paper, we have for the first time broadened this LOC-type analysis to include low ionization AGN, where the spectra become very similar to those of SF galaxies. Diagnostic diagrams that include weak lines can then provide valuable consistency checks for highly ionized AGN when developing physically meaningful interpretations of the AGN locus.

The MFICA technique has identified a statistically relevant set of NLR galaxies that lie along a single curve, which we call the "AGN locus", threading through a 5-dimensional parameter space. An ideal interpretation would be to find a single, tunable physical parameter which recreates the variation from high to low ionization along the AGN locus (first index *i* designating the a*ij* subsets). As we discuss further below, we have found that variations in the radial weighting in the LOC model can account for the differences in the high and intermediate ionization galaxies along the AGN locus in the context of them all being photoionized by similar AGN-like continuum sources.

We note that we used a rather coarse interval of radial and density weightings to match the AGN observations so that their impact on the LOC integrations would be readily apparent from looking at line ratio diagrams. A finer spacing would yield slightly better fitting to all of the observations along the length of the AGN locus, but should not change any of our conclusions.

*5.1.1 Higher – Moderate Ionization AGN*

Fig. 10-14 show that the higher ionization AGN are matched to within a factor of 2 by $\beta = -1.4$ models with either $\gamma = -0.75$ or $\gamma = -0.5$. The only emission-line ratios completely contrary to this statement are [O III] $\lambda 4364$ / [O III] $\lambda 5007$ and [O II] $\lambda 7325$/ H$\alpha$ but even these ratios are in agreement to within a factor of 3 or 4. The ratio [Ar IV] $\lambda 4711$/ [Ar III] $\lambda 7135$ agrees with highest ionization locus point and [O II] $\lambda 7325$/ [O I] $\lambda 6300$ is approximately within a factor 2 of the highest ionization subsample. Although they do not agree for sequentially lower ionization observations, they again are within a factor of 3 or 4. The moderate ionization AGN, which correspond to the middle point along the AGN locus, are best fit by $\gamma = -0.5$ and $\beta = -1.4$. Indeed, every strong-line ratio in our sample agrees with this set of free parameters except for He I $\lambda 5876$/ H$\beta$. In addition to fitting ratios with strong lines, we have also incorporated many weak lines as consistency checks, and these also fit the higher to moderate ionization spectra. These results indicate that the progression along the AGN locus from high ionization to moderate ionization can be represented by a positive change in the radial weighting. Indeed, Figs. 8-12 show this trend to be true for the majority of line-ratio diagrams presented. The change in radial weighting is in the sense that the galaxies with more luminous AGN (the ones higher up the AGN sequence) have more centrally concentrated NLRs.

However, there are still obvious discrepancies in [O III] $\lambda 4363$/ [O III] $\lambda 5007$, [N II] $\lambda 5755$/ $\lambda 6584$ and [Ar IV] $\lambda 4711$/ [Ar III] $\lambda 7135$ for which we do not have an explanation. In particular, the [S II] $\lambda 6716$/ [S II] $\lambda 6731$ and [O III] $\lambda 4363$/ [O III] $\lambda 5007$ ratios indicate low density, high ionization gas defining the so called "temperature problem" in photoionization models (Komossa & Schulz 1997). As in F97, this problem has been solved for higher ionization AGN but there does not seem to be a relationship between a decrease in ionization and our free parameters. This is discussed in more detail below.

*5.1.2 Low Ionization AGN*

Several line ratios are successfully fitted over the entire length of the AGN sequence including the low-ionization cases. Specifically, [O III] $\lambda 5007$/ H$\beta$, [N II] $\lambda 6584$/ H$\alpha$, [O I] $\lambda 6300$/ [O III] $\lambda 5007$, [S II] $\lambda 6716$/ [S II] $\lambda 6731$, He II $\lambda 4686$/ He I $\lambda 5876$, [N I] $\lambda 5200$/ [N II] $\lambda 6584$, fit every observational point over the full range in ionization, and [O III] $\lambda 5007$/ [Ar III] $\lambda 7135$ and [N I] $\lambda 5200$/ H$\alpha$ are very nearly within a factor of 2 for every subsample. However, there are still many line ratios that are not fit at low ionization. In addition to the [O III] $\lambda 4363$/ [O III] $\lambda 5007$, [N II] $\lambda 5755$/ $\lambda 6584$ and [Ar IV] $\lambda 4711$/ [Ar III] $\lambda 7135$ ratios mentioned in the previous section, [O I] λ6300/ Hα also deviates from the trend that a more positive radial weighting corresponds to a lower ionization point on the locus.

We chose to vary the SED, metallicity, integration limits, density weighting and radial weighting, as these were the most likely parameters to provide a simple understanding of the AGN locus. Not counting the integration limits, we have used four free parameters to fit 24 different line ratios. The bottom rows in Tables 3 and 4 show the fraction of line ratios that are fitted to within a factor of two by our models. We fitted about 80 per cent of the observed ratios at the higher ionization end of the AGN sequence, but this drops to about 50 per cent (11–13 ratios fitted) in the two lowest-ionization AGN subsets. The variation in the free parameters cannot completely account for the lowest ionization AGN, but varying other parameters could possibly produce an explanation still within the context of a photoionized, LOC-type model. The simulations in this paper ended when temperatures fell outside the range in which optical emission is strong. Other stopping conditions, such as cloud thickness, could be implemented to test their effects. We have also assumed a simple constant density model for our simulations. Constant density and constant pressure conditions have been implemented in other models (i.e. G04a, G04b) and using similar assumptions in our individual cloud models could produce better fits at the low ionization end of the AGN locus.

Another consideration is whether photoionization or shocks are the prevailing excitation mechanism for low ionization AGN. It is generally accepted that photoionization is responsible for exciting the NLR in luminous Seyfert galaxies. Early work suggested that shock heating was applicable in LINERS or radio galaxies (Dopita & Sutherland 1995, 1996) but this fails to reproduce the emission found in Seyferts and our lowest-ionization subsets. Shock models are not likely to explain the entire AGN locus.

*5.1.3 Interpreting the Radial Distribution of Clouds*

We studied the relationship between AGN spanning a wide range of ionization. We found that the radial distribution of their NLR clouds, or more specifically, the concentration of the clouds towards the central ionizing continuum source, can explain the differing properties of these AGN. Higher ionization AGN, such as Seyferts, have more centrally concentrated gas, while lower ionization "transition" AGN have their NLR spread out over more extended regions of the host galaxies. We are unable to comment on the nature of LINERs since our AGN subsets narrowly miss this region of the BPT diagram (Fig. 3), and clearly miss the region of LINERs identified by Kewley et al. (2006) on other diagrams including [S II]/Hα and [O I]/Hα. The above discussion is in terms of our results for dust-free models, but the dusty models lead to the same conclusions.

Table 1 shows that $L$([O III]) increases fairly strongly as the ionization level increases. However, this is just an ionization effect caused by the clouds being able on average to more efficiently produce [O III] emission. Kauffmann et al. (2003) found in their sample of SDSS AGN that $L$([O III]) is fairly constant as a function of distance from the star-forming locus, which they interpreted as support for the idea that in their sample the increase in ionization level at greater distances from the star-forming sequence represents a "mixing sequence" involving ionization both by starlight and by non-thermal AGN emission. Here we have found that $L$([O III]) does systematically depend on the position along the MFICA AGN sequence, which is roughly equivalent to the distance from the SF locus as measured by Kauffmann et al. (2003). We view this as an encouraging sign that we have cleaned up the Kauffmann et al. (2003) sample to produce one more dominated by just a single type of ionization and therefore better suited to studying the effects of additional variable parameters such as the size of the NLR.

A more fundamental result from Table 1 is that although the observed $L$(Hβ) is roughly constant along the entire length of the AGN sequence, the dereddened $L$(Hβ) values are higher for the lower-ionization objects and then decrease by a factor of about 5 between the low-ionization end of the AGN sequence and the high-ionization end. Since the NLR is in photoionization equilibrium, $L$(Hβ) ∝ $L_{ion}$, assuming that the covering factor and any extinction of the ionizing continuum as seen by the NLR clouds is the same for all AGN subsets. Therefore, we can infer from $L$(Hβ) that the ionizing luminosity drops as the ionization level increases along the AGN locus. We could not measure the ionizing luminosity more directly, since even the high ionization subsets do not show any evidence of a non-thermal continuum, and X-ray measurements have not been made for the AGN in our subsets.

Fig. 15 shows that as $\gamma$ increases, corresponding to a more extended NLR, $r_{char}$ for Hβ and [O III] λ5007 also increases to larger radii due to the equivalent widths at greater distances from the source being more heavily weighted. The maximum $r_{char}$(Hβ) is for the lowest ionization subset (a01) and is 7 kpc, about the same as the SDSS fiber size (with the caveat that the maximum radius $r_{max}$ included within this particular LOC model would be about 1.5 times the SDSS fiber size). This is consistent with typical observed sizes of resolved NLRs (e.g. Bennert et al 2002; 2006a; 2006b) Our dusty models predict the same trends for $r_{char}$(Hβ) and $r_{char}$([O III]) with ionization and predict similar sizes as well.

We deduced that the underlying physical trend is that the objects with more luminous central continuum sources have more extended NLRs. A possible explanation is that the NLR might be a wind driven by radiation pressure. The decrease in ionization level in the more luminous objects is just an incidental result of their NLRs being more extended. The change in the radius weighting might represent a "clearing out" of the central regions as the AGN starts influencing the structure of the NLR. The usual scenario is that the AGN epoch begins after the ULIRG epoch and results in winds. The change in radius weighting from central to more distant might represent blowing out material from closer in, leaving distant material in place.

*5.1.4 Is the $L_{ion}$ trend real?*

This trend in the inferred $L_{ion}$ is a surprising result. Previous studies have assumed that the [O III] λ5007 luminosity is a direct indicator of the AGN bolometric luminosity (e.g. Kauffmann & Heckman 2009; Heckman et al. 2004). [O III] λ5007 is used because it suffers less contamination from SF than other strong lines in SDSS spectra or Hβ. In contradiction to these previous studies, our dereddened $L^c$([O III] λ5007) values decrease with increasing $L_{ion}$, the opposite of the trend in $L$(Hβ). These earlier studies typically involved samples rather similar to our parent SDSS sample, but the AGN locus picks out from that larger sample a class of objects in which the ionization level is anticorrelated with $L_{ion}$ and this effect has overwhelmed the underlying luminosity correlations.

Along the same lines, Lamastra et al. (2009) find a clear positive correlation between $L^c$([O III]) and the X-ray luminosity $L_x$. To the extent that $L_x$ is a measure of $L_{ion}$, this again disagrees with our result. However, Vasudevan & Fabian (2006) showed that the relationship between X-ray bolometric correction factor and luminosity is not at all obvious, leaving the interpretation of the relationship between $L(\lambda 5007)$ and $L_{ion}$ unclear. Furthermore, the positive correlation between $L^c$([O III]) and $L_x$ in Lamastra et al. (2009) spans 6 orders of magnitude, while the values of $L^c$[O III] for the AGN subsets differ only by a factor of 4. Indeed, in the narrow range of dereddened [O III] luminosities we have isolated (~$10^{41.5}$ erg s$^{-1}$), their sample does not indicate a clear linear trend. Additionally, the Lamastra et al. (2009) sample was over a larger redshift range, $0.3 < z < 0.8$, than the sample considered in this paper.

While Seyfert galaxies with X-ray observations do not show the same trends that we show here, observations of galaxies extending down into the composite region of the BPT have shown these correlations. Zhang et al. (2013) used a similar sample of narrow-line dominated low-$z$ ($0.02 > z > 0.32$) galaxies from SDSS. They found that Seyferts have slightly larger [O III] luminosities compared to composites, however Seyferts also have approximately a factor of 5 smaller H$\beta$ luminosities than composites. Our sample does not contain composite galaxies, which have a mixture of AGN and SF contributions, but our low ionization AGN do fall into the composite region of the BPT diagram.

The $E(B-V)$ values indicate that the low-ionization objects (the more extended NLRs) are more strongly reddened than are the high-ionization objects, and the luminosity correlation does not appear until the reddening correction has been applied. Could the luminosity correlation could be an artifact due to an incorrect reddening correction? We used H$\alpha$/H$\beta$ as our primary reddening indicator and assumed that the lines have their Case B intensities. Cloudy explicitly solves a many-level model of H I emission, including radiative transfer and collision processes (Luridiana et al. 2009), so our calculations result in explicit predictions of the H$\alpha$/H$\beta$ ratio. These are listed in Table 3. While the predicted Balmer decrements listed in Table 3 and 4 are slightly steeper than the Case B value that we used (log H$\alpha$/H$\beta$ = 0.46), there is little variation in the predicted H$\alpha$/H$\beta$ ratio, so that the physics of the emission line formation could not create the effect we observe.

The reddening correction to $L$(H$\beta$) using the H$\alpha$/H$\beta$ ratio has little dependence on the exact reddening law used. Using the Cardelli et al. (1989) curve with $R_v$ = 3.1 instead of $R_v$ = 2.86 gives little variation (~6%), as does the reddening law in Wild et al. (2011), with a maximum deviation of ~30%. We have assumed that $L_{ion} \propto L^c$(H$\beta$), which could be incorrect if low-ionization objects (ones with larger $L$(H$\beta$)) view the continuum source through a different amount of UV extinction than high-ionization, lower $L$(H$\beta$) objects. For this to be the source of the inferred trend in $L_{ion}$, the high-ionization objects would need to see the continuum source through a *larger* amount of extinction than the low-ionization objects. Similarly, applying an attenuation curve (Calzetti, Kinney & Storchi-Bergmann 1994; Calzetti et al. 2000) creates for need for even larger $L_{ion}$ at low ionization due to greater extinction. Therefore, we conclude that there are ways that our inferred inverse correlation between $L_{ion}$ and ionization level could be incorrect; however, there is no obvious evidence for such effects in our sample taken on its own.

Possible selection effects in the SDSS sample must also be considered. The basic SDSS galaxy sample is continuum-flux limited. As is described in Paper I, we selected emission-line galaxies from that sample using criteria that include a S/N ratio limit on the H$\beta$ emission line and a very narrow redshift range, so our AGN sample is effectively H$\beta$-luminosity limited. For this reason, it is not surprising that the observed $L$(H$\beta$) given in Table 1 is constant across all of the subsamples. The strong trend in $L^c$(H$\beta$) therefore comes strictly from the observed trend in the Hα/Hβ ratio from which we deduce E(B-V). Can this Hα/Hβ trend be due to selection effects? For this to be true, we would need to either be systematically missing the high-ionization galaxies with the strongest Hα lines or the low-ionization galaxies with the weakest Hα lines. The latter seems more probable, but even for no reddening the Hα line is three times stronger than Hβ, so we do not see why such lines would be missed or incorrectly measured.

The actual subsamples used here reflect the AGN locus that was defined by the MFICA techniques described in Paper I. We are confident that we have correctly studied the physical properties of that sample.

*5.1.5 Luminosity correlations*

Our dust free models also predict that $L_{ion}\Omega$ is anti-correlated with ionization (Table 3), but our dusty models predict essentially no correlation of $L_{ion}\Omega$ with ionization (Table 4). Since our dusty models are not firmly ruled out, this presents two different scenarios about whether the ionizing luminosity changes as function of ionization. A constant $L_{ion}$ is the simplest case, but our observations show that $L^c(H\beta)$ and $L^c([O\ III])$ vary with ionization and so the real physical picture is likely to be complicated in nature.

These results also relate to the bolometric correction factor $C_{OIII} = L_{bol}/L^c([O\ III])$, where $L_{bol}$ is the bolometric luminosity. Lamastra et al. (2009) found $C_{OIII}$ to exhibit a positive correlation with $L^c([O\ III])$ in their data set. Since $L_{ion} \propto L_{bol}$, we can determine the trend in bolometric correction across our AGN subsets. While $L^c([O\ III])$ is typically used to estimate the ionizing luminosity, the [O III] line is subject to multiple ionization effects in contrast to $L^c(H\beta)$ which simply counts photons due to the balance between photoionization and recombination. The weakness of Hβ in combination with absorption from the galaxy prevents this line from being commonly used, however our relatively clean continuum-subtracted spectra allow us to probe the physics of the NLR more deeply than would be possible by only using $L^c([O\ III])$.

Tables 3 and 4 list $L_{ion}\Omega/L^c([O\ III])$ and $L_{ion}\Omega/L^c(H\beta)$, which are proportional to the bolometric correction if $\Omega$ is constant. For our models, the bolometric correction determined from $L^c([O\ III])$ decreases as one moves to higher ionization subsets. In contrast, the sample presented by Lamastra et al. (2009) shows a positive correlation between $C_{OIII}$ and $L^c([O\ III])$. But given the small range of luminosities considered in our sample and the much larger range considered in Lamastra et al. (2009), their finding is reasonably consistent with our results.

For the dust-free models, $L_{ion}$ increases by a factor of 6 as one moves towards lower ionization subsets and $L^c(H\beta)$ correspondingly increases by the same amount. For a constant covering factor, this simply argues that more ionizing photons increases the depth at which the ionization front forms resulting in stronger $L^c(H\beta)$. The situation is more complex for the dusty models.

## 5.2 [O III] and [N II] Temperatures

The major problem with these LOC models is that they do not reproduce the increase in the temperature-sensitive [O III] λ4363/λ5007 and [N II] λ5755/λ6584 ratios that is observed as one moves down the AGN sequence towards lower ionization (Fig. 12), nor do the dust-free models reproduce the very high ratios that are reached. Our large galaxy sample has allowed this effect to be seen not only for the typical [O III] ratio but also for the [N II] ratio. Although neither ratio can be measured in the very lowest ionization case, Fig. 4 shows that [O III] λ4363 is easily measured for all other cases and likewise, [N II] λ5755 is easily measured in our high – moderate ionization subsets. In the low-density limit, the observed [O III] ratio for the high-ionization end of the AGN sequence gives a temperature of ~11,000 K and the observed [N II] ratio gives a temperature of ~9,000 K. The lowest-ionization subset for which the [O III] ratio can be measured (the a11 case) corresponds to a temperature of ~20,000 K in the low density limit. High values for these auroral/nebular line intensity ratios have long been known to exist in some AGN (Shuder & Osterbrock 1981) and have also been discussed by Komossa & Schulz (1997) and Zhang, Liang & Hammer (2013).

The [O III] λ5007 and [N II] λ6584 lines become collisionally quenched at high densities, while [N II] λ5755 and [O III] λ4363 optimally emit. This could mean that the so called "temperature problem" is actually a density problem and that the [O III] and [N II] ratios are giving false temperatures. We used simple power laws to represent the radial and density distributions in our LOC models, but there could in principal be some other arrangement of the NLR that emphasizes high density gas. However, there are several problems with this scenario.

First, the [N II] λ5755 and [O III] λ4363 optimally emit in different regions of the LOC plane. So emphasizing a moderately high density (e.g. $\log(n_H) \sim 6.5$) would solve the [N II] temperature problem but give an even worse fit to the [O III] temperature. Second, our low-density indicating line ratio, [S II] λ6716/ [S II] λ6731, fits the entire AGN sequence to within ~10 per cent and our high-density indicating line ratio, [S II] λ4070/ [S II] λ6720, fits the AGN sequence to within a factor of 2 or 3 down to all but the lowest ionization observed AGN subset. Finally, additional high density components would add substantial Balmer emission (since these lines emit strongly over a wide density range) which would result in underpredicting many key ratios (e.g. [O III] λ5007/ H*β*, [S II] λ6720/ Hα, etc.). Indeed, Komossa & Schulz (1997) attempted to solve this temperature problem by emphasizing a high density component, only to find that it severely overestimated [O I] emission. Therefore, it is likely that the low ionization AGN really do contain low density gas with high $T_e$.

Fig. 9 shows that [O III] λ4363 optimally emits at $\log(r) \sim 18.0$ while [O III] λ5007 optimally emits at $\log(r) \sim 19$. Similarly, [N II] λ5755 optimally emits at $\log(r) \sim 19.5$ while [N II] λ6584 optimally emits at $\log(r) \sim 21.0$. Therefore smaller values of the radial exponent γ (which correspond to less extended NLRs) should give larger [O III] and [N II] ratios and hence larger deduced $T_e$. However, the observations surprisingly show that the higher ionization cases have smaller [O III] and [N II] ratios even though the other line ratios indicate that higher ionization cases are the ones with the more concentrated NLRs (smaller fitted values of γ).

One possible explanation for the high temperatures is an increasing contribution of shock excitation at lower ionization. The shock models presented by Dopita & Sutherland (1995) do reach the high [O III] temperatures found in our subsets. However, their shock models that produce these high temperatures significantly overestimate the [O I] emission given in our lower ionization subsets. Most importantly, their shock models *cannot* reproduce the emission found in high excitation Seyfert galaxies, so our hope of unifying the low- and high-ionization AGN would have to be abandoned.

Another factor that could explain the high [O III] ratio in our low-ionization AGN is a decrease in metallicity. A decrease in metal abundances results in a decrease in cooling which raises the electron temperature. Komossa & Schulz (1997) studied this effect in luminous Seyfert 2 galaxies. Their models that resulted from scaling solar metal abundances by 0.3, which they favored, reproduce the [O III] ratio found in our low ionization subsets. This effect has an upper limit of log([O III] $\lambda 4363/ \lambda 5007$) ~ -1.3, but this is already larger than any [O III] ratio found in our subsets. However, studies with far larger samples than the one used by Komossa & Schulz (1997) have shown that Seyferts with subsolar metallicities are rare (Groves et al. 2006), so this explanation may be limited to low ionization AGN.

Photoelectric heating from grains can also increase the electron temperature enough to reproduce the temperature sensitive ratios found in the high-ionization Seyfert galaxies studied by Komossa & Schulz (1997) and G04b. Our dusty models have the same effect, producing an [O III] $\lambda 4363/ \lambda 5007$ ratio that matches the moderate ionization subsets, but they do not accurately predict the highest ionization observations.

This problem might be ameliorated in models adding cosmic rays, shock excitation, turbulent heating or an increase in grain abundance in more extended regions of the NLR. Although a full study must be left for a future paper, we briefly explored one of these possibilities by running our dust-free and dusty model with a cosmic ray ionization rate $10^3$ times greater than the Galactic background given by Indriolo et al. (2007). It is possible that AGN and starburst galaxies see a greater cosmic ray flux than other galaxies (Papadopoulos 2010). However, the extra heating by cosmic rays had a negligible effect on the [O III] and [N II] temperatures in both our dust free and dusty models, leaving the need for a deeper investigation of an additional component to address this "temperature problem" in low ionization AGN.

### 5.3 The Possibility of a Dusty NLR

We were unable to find a set of free parameters for our dusty models that can account for systematic variations along the AGN locus. However, our results do not rule out the dusty models. Describing the NLR as either completely dusty or dust-free is likely to be a gross simplification for a complicated environment; it is indeed plausible that both dusty and dust-free regions exist.

A dusty, radiation pressure dominated NLR has previously been proposed by G04a and provides a different interpretation of the physical meaning behind the AGN locus. The basic premise behind that model is that radiation pressure exerted by the ionizing source upon grains, the photoelectric heating produced by the grains, and the ionizing photon absorption are non-negligible effects in the NLR. Similar effects occur in individual Galactic H II regions (e.g. the Orion Nebula; Baldwin et al. 1991). Aside from these effects, grains also deplete refractory elements and destroy resonance lines (e.g. Ly$\alpha$ $\lambda 1216$).

In this case, the theoretical diagnostic diagrams computed by G04b indicate that for Z = 2Z_sun a decrease in ionization along the AGN sequence represents a decrease in ionization parameter for a dusty, constant pressure NLR. The G04a models do not even consider the low ionization end of the AGN locus, nor do they fit the observed values for those cases. In addition, their interpretation was mostly limited to strong lines. For the weak line ratios that were included in this analysis, our dust-free models fit our subsets better than our dusty models and their dusty models clearly do not reproduce our subsets. In fact, their constant density, dust-free models are in better agreement with these lines.

Dusty models, as in G04b for example, are only capable of reproducing a narrow range of high ionization UV lines in NLR Seyferts (Nagao, Maiolino & Marconi 2006). In particular, dusty models for C IV λ1551/ He II λ1640 > 1.5 fail to match observations due to stagnation at low metallicity and high ionization parameter, while the dust-free models from F97 are successful. This suggests that at least part of the high ionization region of the NLR is dust free. One current physical picture envisions the NLR as a mixture of a dusty region responsible for lower ionization emission (e.g. [O III], [S II]) and a dust free component that emits coronal lines (e.g. [S VIII], [Fe X]) (Dopita et al. 2002).

Our analysis results in contradictory conclusions about the dust content. Komossa & Schulz (1997) favored a dust-free model mainly because their dusty model failed to predict strong, moderate ionization Fe line intensities (e.g. [Fe III] λ4658, [Fe VII] λ6087) due to highly depleted Fe abundances. F97 found that lower ionization regions were dusty and higher ionization regions dust free.

Low ionization lines may suggest that the gas is dusty. The [Fe III] line is not detected, with I([Fe III] λ4658)/I(Hβ) < 0.02–0.04. The predicted [Fe III]/Hβ ratios range from 0.05 for our highest ionization dust-free LOC model to 0.01 for our moderate ionization dusty model, which are too close to the observational limits to be conclusive. But at most, marginal depletion of Fe might be required to fit the observations. The [Ca II] λλ7291, 7324 doublet is typically strong in dust-free regions (Kingdon, Ferland & Feibelman 1995). The [Ca II] λ7324 line is blended with [O II] λ7325, but we do not detect the [Ca II] λ7291 line. Our dust-free models predict that I(λ7291) / I(Hβ) = 0.31 for higher ionization AGN, and they also overpredict the [Ca II] λ7324/ Hβ ratio at I(λ7324) / I(Hβ) = 0.21 since it is predicted to be larger than the [O II] λ7325/ Hα emission line ratio. This indicates significant depletion of Ca onto dust.

On the other hand, high ionization lines suggest that the gas is dust free. The [Fe VII] λ6087 emission line is observed to be moderately strong in our higher-ionization subsets. The dust free model does correctly predict the [Fe VII] λ6087/ Hα ratio over the range of ionization that it is observed, while the dusty model underpredicts this line by almost a factor of three.

All of this leaves the overall picture of the NLR's dust content unclear, and it likely that both high ionization dust-free and lower ionization dusty clouds exist in regions of the NLR. An ablating wind, as mentioned above, from inner regions could be dust-free and lead to a mixture of dust abundances.

F97 came across a similar problem. They found that low-ionization regions appeared to be dusty while high ionization region must be dust free to create strong high ionization Fe lines. The spectral diagnostics do not clearly indicate the ionization potential where the change occurs, so we cannot say whether any particular line might form in a dusty or dust free region. The results in Fig. 16, which show that the intensity ratios involving low-ionization lines, such as I([N I] λ5200)/I([O II] λ7325) or I([O I] 6300)/I(Hα), fit the dust-free models better than the dusty models add to the confusion.

### 5.4 Is the AGN Sequence Actually a Mixing Sequence?

The AGN and SF sequences defined by MFICA are both *necessary* and *sufficient* to describe the spectra of all the galaxies in the sample. Thus, it would take a rather contrived set of circumstances to interpret these sequences as varying mixtures of AGN and SF contributions.

However, when plotted on a BPT diagram, the AGN sequence described here bears some similarity to a mixing sequence of the sort used by Kauffmann & Heckman (2009). This interpretation, based purely on empirical modeling, assumes that moving down the sequence to lower ionization galaxies is the result of an increasing SF component. However, the AGN locus was constructed in such a way as to isolate pure AGN, i.e. those that are not mixed with any significant emission from star formation. The purity can be seen more clearly in the five-dimensional space of MFICA weights (see Paper I). For example, Paper I shows that $W_4$ is elevated all the way along the AGN locus, a feature inconsistent with mixing.

At the high-ionization end of the AGN locus, Fig. 3 and Table 1 show that a41 has essentially zero contribution from components 2 and 3. At least one of these components is present throughout the star formation locus (Fig. 3c), so subtracting a star formation contribution from a41 would lead to negative weights in one or both of the components. But negative weights correspond to the non-physical case of negative emission and hence can be ruled out, meaning that a41 must represent a pure non-SF spectrum. Further down the AGN sequence there are non-zero contributions from component 3, so it would be possible to subtract some high-metallicity star formation and still be left with a valid (i.e. non-negative) set of MFICA weights.

To investigate this line of reasoning, we checked to see whether a set of mixing models could accurately describe the AGN sequence. We mixed the pure AGN subset a41 with the lowest ionization pure SF subset s01 so that the spectra in these exactly reproduced the [O III] / H$\beta$ in the remaining AGN subsets. The sums of the AGN and SF contributions were forced to equal unity. We note that we did not also attempt to exactly reproduce the [N II] / H$\alpha$ ratio because this ratio is essentially constant throughout all of the a$i$1 spectra.

The resulting contributions and emission line ratios for these mixing models are shown in Table 5. The a41 subset perfectly matches the observations, by definition, and then the contribution of AGN gets nearly cut in half as one proceeds to the lower ionization subsets. The lowest ionization galaxies are almost completely dominated by SF. Our resulting models fit the observations very well as indicated by the number of fitted line ratios given the last row. Although, there are several caveats to keep in mind. As mentioned previously, the models fit the a41 perfectly due to the way we have developed the models. The low S/N in the s01 subset has caused many of the predictions to be upper limits. Finally, [O III] / H$\beta$ also should fit every subset by definition. Even with these exceptions however, the mixing models fit the vast majority of line ratios within a factor of two.

So, is the AGN sequence a sequence of pure AGN or a mixing sequence? Table 5 shows that the mixing model fits almost all of the observed line ratios to within a factor of two. However, the MFICA components contain more information than just the line intensity ratios that we have analyzed so far. The following discussion takes advantage of that additional information, although it does not allow for uncertainties in the placement of the SF and AGN loci. A number of lines of evidence suggest that the locus does still represent a pure AGN sequence.

First, the AGN locus is both *sufficient* and *necessary* to describe the observed data points, when combined with the SF locus. It is sufficient because each data point in the input sample can be described by mixing a point inside the AGN locus with one inside the SF locus. It is necessary because if the locus is shortened at either end, a population of data points is revealed that can no longer be described by such a mixture. In particular, if just the a41 point replaces the AGN locus, then it is impossible to reproduce many of the observed points from a linear combination of SF and AGN spectra.

Second, in spite of our factor-of-two success in Table 5, we cannot really replicate the AGN locus by a linear combination of any single point along the SF locus and a41. This can be seen on Fig.3; the line joining a41 and s01 falls on the AGN sequence in panels (a) and (c), but not in panel (b). If instead we used the point on the SF locus that has the same $W_3$ value as a01, then a line joining that SF point to a41 falls on top of the AGN locus on panel (b), but not on panels (a) or (c). Nor can any other combination of a single SF point and a41 reproduce all points along the AGN locus.

Third, the successful fitting of the observations with a mixing sequence is partially expected given the purely empirical nature of the mixing models. The empirical modeling masks the inherent physical complexity of these objects. For example, the temperature problem discussed in §4.2 has been "solved" by the mixing sequence in the sense that it is just an observed fact, however the mixing models provide no physical reason for why having a large fraction of SF photoionization would produce such high temperatures. If instead of using the observed spectra we had tried mixing our best-fitting photoionized models for a41 and s01, the fits at all locations along the sequence (including at a41) would not have been nearly as good.

Finally, we interpret the observed galaxies that lie between the SF and AGN sequences in Fig 3 to be the true composite cases, because their spectra can be reconstructed from some mix of a point along the SF sequence with some point along the AGN sequence. On the BPT diagram (Fig. 1), the SF locus demarks the extreme lower-left edge of the arc of galaxies in the SF region, while the AGN locus is close to the right-hand edge of the observed AGN population, so the galaxies between these two loci are obvious candidates for SF-AGN composites based on this 2-dimensional parameter space. However, there are additional objects which fall on top of the AGN locus when projected onto the BPT diagram, but which are not on the AGN locus when mapped in the 5-dimensional space of the MFICA components. This indicates a real difference in their emission-line spectra, the most obvious of which are that for the same [O III]/Hβ ratio a subsample of galaxies which the MFICA analysis identifies as having SF contributions greater than 20% have smaller observed Balmer decrements and, irrespective of reddening corrections, smaller I([Ne III] 3869)/I([O II] 3727 ratios than galaxies in the same position on the BPT diagram that lie on the AGN sequence. This difference could in principal be picked out using additional line ratio diagrams, but it is more obvious using the MFICA technique.

Taken together, these lines of evidence limit the possible interpretations to three options:

(1) The AGN locus represents spectra with no significant contribution from SF.

(2) The AGN locus represents mixtures of the a41 spectrum with points from the SF locus, but the SF spectrum in question varies along the length of the AGN locus, and is constrained in the proportions in which it mixes with the a41 spectrum.

(3) The AGN locus represents mixtures of low-ionization SF with a range of non-SF spectra, but each non-SF spectrum is always mixed with a specific proportion of SF. The proportion of SF falls to zero as the non-SF spectrum approaches a41.

We further note that the constrained mixing proportions in (2) and (3) must survive despite the factor of 1.4 difference in aperture between the highest and lowest redshift observations in the input sample, given the fixed size of the SDSS fibers. While we cannot completely eliminate options (2) and (3), the fine tuning of the observed mixtures that they require, together with our success in modeling most of the AGN locus with pure AGN models, lead us to conclude that the AGN locus does not include any significant contributions from SF.

A future follow-up study is needed to compare the optical morphology and stellar contribution from the galaxies to their positions along the AGN sequence. Our MFICA component fitting to the continuum in principal already provides a quantitative description of the stellar continuum spectrum and morphological classifications have been performed by, for example, the Galaxy Zoo project (Lintott et al. 2011). Another avenue for future work is to construct additional tests of whether the spatial concentration of the NLR truly is a key parameter for determining its ionization level. One approach would be to repeat this study for a sample at a closer distance where the SDSS fiber captures a smaller fraction of the host galaxy, biasing the results towards more concentrated NLRs, to see if this changes the distribution of galaxies in the BPT and MFICA space. Another approach would be to obtain long slit or IFU spectra of resolved objects and directly measure the spatial concentration of the NLR. We are indebted to the referee for suggesting these additional lines of investigation.

## 6. Conclusions

The MFICA analysis described in Paper I uncovered a statistically meaningful sequence of narrow- emission-line galaxies that lie along an AGN locus and which we argue are distinct from galaxies with NLR's powered by starbursts. We have used LOC models to investigate what fundamental variable determines a galaxy's position along this single-parameter AGN sequence.

We have found that LOC models in which the ionizing continuum is always from an AGN and in which only the radial extent of the gas is varied successfully reproduce most of the BPT and VO87 diagnostic line ratios observed over the sequence of high- through moderate-ionization NLRs in narrow-lined AGN. This includes fitting the line ratios which depend most strongly on the abundances, gas density and spectral energy distribution. Our sequence of models also fit a number of the observed line ratios even for very low-ionization AGN. The most significant discrepancies between the observations and our models are that we cannot fit several important line ratios in the low-ionization objects, and in particular the models do not reproduce the high $T_e$ implied by the large observed [O III] $\lambda4363/\lambda5007$ ratio.

Spectral signatures of the dust content are inconclusive, as found in previous studies. It is likely that different regions have a different dust to gas ratio, perhaps because of their history. Luckily, the key spectral diagnostics used to reach the main conclusions of this paper are insensitive to whether or not our LOC models contain dust.

We have shown that the high - moderate ionization objects along the AGN locus generated by MFICA represent a physically meaningful sequence in NLR properties. The most viable interpretation describes the sequence from high ionization to low ionization AGN as a change in the radial distribution of the NLR gas, which is proportional to changing the flux that is incident on each cloud. This finding states that higher ionization AGN contain optimally emitting clouds that are more concentrated towards the central continuum source than in lower ionization AGN, but that the density distribution is on average the same. According to the LOC models, the characteristic physical sizes of the NLR as measured in Hβ and [O III] are much larger for the low-ionization (high luminosity) objects than for the high-ionization (low luminosity) objects. Scaling from the Hβ luminosity, we also found that the central continuum source in low-ionization objects is more luminous than in high-ionization objects. This suggests that the ionization sequence might be a sequence in how effectively AGN have cleared out their central regions due to radiation pressure, possibly due to a difference in age.

To verify our conclusions pertaining to ionization being anti-correlated with concentration, there are two paths forward, which could be the subject of future work. First, closer samples in SDSS should provide a different distribution of galaxies on BPT diagram due to a bias towards more concentrated objects. Galaxies at lower redshift would allow the SDSS fiber size to cover the entire radial extent of the NLR in low ionization AGN. Secondly, spatially-resolved objects would allow direct confirmation of how NLR concentration varies with ionization.

We compared the results of these LOC models, which attempt a real physical description of an entire NLR, to the results obtained by fitting the AGN sequence with simple empirical combinations of SF and high-ionization AGN spectra. The empirical mixing curves describe the spectra at intermediate points along the AGN locus better than our LOC models do, but this is at least partly because the empirical approach does not require us to actually produce physically self-consistent models of anything. Our basic argument is that a mixing model approach employing just a single AGN template spectrum, as used here, cannot reproduce the distribution of points seen in the MFICA. In that light, we present our LOC models as an alternative that can describe all of the data.

We arrived at these results by not only matching typical line ratio diagnostics based on the strongest emission lines but also by taking advantage of weaker lines for consistency checks. This allowed greater confidence in the temperature, density, abundances and ionizing continuum constrained from strong line ratios. We will continue this analysis with the SF sequence in a future paper where the physical properties of low ionization SF galaxies might overlap with those of low ionization AGN.

## **ACKNOWLEDGEMENTS**

We are grateful to Brent Groves, our referee, for a number of very helpful comments and suggestions. CTR and JAB acknowledge support from NASA ADP grant NNX10AD05G and NSF grant AST-1006593. CTR wishes to acknowledge the support of the Michigan State University High Performance Computing Center and the Institute for Cyber Enabled Research. This work also used the Extreme Science and Engineering Discovery Environment (XSEDE), which is supported by National Science Foundation grant number OCI-1053575. JTA acknowledges the award of an ARC Super Science Fellowship. PCH acknowledges support from the STFC-funded Galaxy Formation and Evolution programme at the Institute of Astronomy. GJF acknowledges support by NSF (1108928; and 1109061), NASA (10-ATP10-0053, 10-ADAP10-0073, and NNX12AH73G), and STScI (HST-AR-12125.01, GO-12560, and HST-GO-12309).

**Appendix A**

Here we present several additional diagnostic diagrams that place weaker constraints, or merely consistency checks, on our models.

Fig. A1 contains indicators of the SED and density. Panel (b) combines two SED indicators, [N I] $\lambda 5200$/ [N II] $\lambda 6584$ and [O II] $\lambda 7325$/ [O I] $\lambda 6300$, both of which are minimally affected by reddening. The results show that although the observations do not show the predicted modest changes of these line ratios as the radial weighting is changed in the LOC model, the entire [N I] $\lambda 5200$/ [N II] $\lambda 6584$ sequence and higher ionization observation of [O II] $\lambda 7325$ / [O I] $\lambda 6300$ still does agree with a change in radial weighting to within approximately a factor of two. Panel (c) on Fig. A1 combines two SED indicating ratios, [S III] $\lambda 9069$/ [S II] $\lambda 6720$ vs. [O II] $\lambda 3727$/ [O I] $\lambda 6300$. The observations used here do not extend out to the [S III] lines, but these lines are often measured in low-$z$ galaxies and we include them here so that other investigations can take of advantage of these results. The ratio [O II] $\lambda 3727$/ [O I] $\lambda 6300$ is included in our observations, and ranges (in the log) from 1.0 at the high ionization end to 1.5 for the lowest ionization cases. The adopted LOC models all predict that this ratio should be about 0.7 in the log, which agrees with the observations to within a factor of two except for the lower ionization cases. Panel (d) combines a low density indicator, [O II] $\lambda 3727$/ $\lambda 7325$, with a high density indicator, [S II] $\lambda 4070$/ $\lambda 6720$ (Holt et al. 2011). Our model agrees with the higher and lower ionization subsamples, but fails to match the moderate ionization AGN subsamples.

Fig. A2 presents abundances diagnostics, several of which have no predicted dependence on $\gamma$ or $\beta$. Panel (a) stems from a discussion in Storchi-Bergmann & Pastoriza (1990) where only nitrogen and sulfur abundances needed enhancements to account for emission in different AGN covering a wide range in ionization. However, our 1.4 Z$_\odot$ abundance set nicely fits this diagram for the high - moderate ionization subsamples. Panel (b) incorporates [O II] $\lambda 7325$/ H$\alpha$, a ratio involving a weak line only measurable because of our large sample, which fits our highest ionization subsamples within a factor of 3. The diagram in Panel (f), [O I] $\lambda 6300$/ H$\alpha$ vs. [N I] $\lambda 5200$/ H$\alpha$ uses lines with lower ionization potentials. In this case the predicted ratios for our adopted sequence of models all cluster at one point on this diagram, and the observed ratios fall at that same location except for the case of [O I] $\lambda 6300$/ H$\alpha$ for the lower ionization subsets.

| Table 1. Properties of the a$ij$ subsets | | | | | | | | | | | | | | | |
|---|---|---|---|---|---|---|---|---|---|---|---|---|---|---|---|
| Subset | a00 | a01 | a02 | a10 | a11 | a12 | a20 | a21 | a20 | a30 | a31 | a32 | a40 | a41 | a42 |
| $E$(B-V) | 0.80 | 0.83 | 0.68 | 0.43 | 0.49 | 0.56 | 0.40 | 0.43 | 0.45 | 0.30 | 0.33 | 0.39 | 0.20 | 0.20 | 0.21 |
| Observed $L$(H$\beta$)[1] | 1.8 | 2.1 | 2.4 | 2.0 | 2.2 | 2.6 | 3.1 | 2.9 | 3.2 | 3.1 | 2.4 | 2.8 | 2.9 | 2.9 | 2.8 |
| Observed $L$([O III] 5007)[1] | 0.95 | 0.85 | 0.92 | 3.1 | 3.6 | 3.7 | 9.2 | 9.6 | 9.1 | 16. | 12. | 13. | 24. | 25. | 24. |
| Observed $L_\lambda(\lambda 5007)$[2] | 0.47 | 0.49 | 0.52 | 0.45 | 0.52 | 0.58 | 0.50 | 0.67 | 0.60 | 0.60 | 0.49 | 0.53 | 0.52 | 0.50 | 0.50 |
| Dereddened $L^c$(H$\beta$)[1] | 29. | 38. | 25. | 9.0 | 12. | 18. | 12. | 13. | 15. | 8.8 | 7.7 | 11. | 5.9 | 5.9 | 5.7 |
| Dereddened $L^c$([O III] 5007)[1] | 14. | 14. | 8.8 | 13. | 18. | 24. | 35. | 40. | 41. | 43. | 35. | 47 | 47. | 49. | 48. |
| Dereddened $L^c_\lambda(\lambda 5007)$[2] | 6.97 | 8.14 | 4.94 | 1.88 | 2.60 | 3.77 | 1.89 | 2.80 | 2.70 | 1.62 | 1.42 | 1.91 | 1.02 | 0.98 | 0.99 |
| $W_\lambda$(H$\beta$) (Å) | 3.81 | 4.25 | 4.64 | 4.47 | 4.23 | 4.48 | 6.23 | 4.32 | 5.35 | 5.13 | 4.94 | 5.30 | 5.55 | 5.78 | 5.65 |
| $W_\lambda$([O III] 5007) (Å) | 2.01 | 1.72 | 1.78 | 6.93 | 6.92 | 6.37 | 18.5 | 14.3 | 15.2 | 26.5 | 24.7 | 24.6 | 45.9 | 49.8 | 48.4 |
| Component weights | | | | | | | | | | | | | | | |
| Component 1 | 0.03 | 0.02 | 0.01 | 0.14 | 0.13 | 0.11 | 0.25 | 0.25 | 0.25 | 0.40 | 0.40 | 0.40 | 0.54 | 0.57 | 0.59 |
| Component 2 | 0.02 | 0.01 | 0.01 | 0.01 | 0.01 | 0.02 | 0.02 | 0.01 | 0.01 | 0.00 | 0.01 | 0.01 | 0.01 | 0.00 | 0.00 |
| Component 3 | 0.73 | 0.74 | 0.75 | 0.52 | 0.54 | 0.55 | 0.34 | 0.34 | 0.34 | 0.13 | 0.16 | 0.18 | 0.01 | 0.01 | 0.01 |
| Component 4 | 0.09 | 0.17 | 0.24 | 0.16 | 0.34 | 0.32 | 0.17 | 0.26 | 0.35 | 0.22 | 0.30 | 0.38 | 0.22 | 0.26 | 0.29 |
| Component 5 | 0.13 | 0.07 | 0.00 | 0.17 | 0.09 | 0.00 | 0.22 | 0.14 | 0.05 | 0.24 | 0.14 | 0.04 | 0.23 | 0.16 | 0.11 |

[1]Emission-line luminosities are in units of $10^{40}$ erg s$^{-1}$.
[2]Continuum luminosities are in units of $10^{40}$ erg s$^{-1}$ Å$^{-1}$.

| Table 2a. Measured emission line strengths for the AGN locus, relative to Hβ. |||||||||||||||||
|---|---|---|---|---|---|---|---|---|---|---|---|---|---|---|---|---|
| Ion | λ | a00 | a01 | a02 | a10 | a11 | a12 | a20 | a21 | a22 | a30 | a31 | a32 | a40 | a41 | a42 |
| [O II] | 3727 | 1.47 | 1.43 | 1.18 | 1.48 | 1.30 | 1.31 | 1.78 | 1.58 | 1.48 | 2.05 | 1.82 | 1.71 | 2.48 | 2.37 | 2.35 |
| H I | 3798 | <0.03 | <0.04 | <0.04 | 0.01 | 0.03 | 0.04 | 0.01 | 0.03 | 0.05 | 0.03 | 0.01 | 0.01 | 0.05 | 0.04 | 0.04 |
| H I | 3835 | <0.01 | <0.02 | <0.05 | 0.02 | 0.02 | 0.03 | 0.01 | 0.03 | 0.04 | 0.03 | 0.01 | 0.02 | 0.08 | 0.07 | 0.08 |
| [Ne III] | 3869 | 0.10 | 0.02 | 0.08 | 0.16 | 0.17 | 0.18 | 0.41 | 0.42 | 0.38 | 0.57 | 0.54 | 0.56 | 0.78 | 0.79 | 0.81 |
| H I | 3889 | 0.04 | 0.03 | 0.08 | 0.11 | 0.09 | 0.11 | 0.14 | 0.14 | 0.14 | 0.13 | 0.13 | 0.14 | 0.16 | 0.16 | 0.16 |
| [S II] | 4070 | 0.12 | 0.06 | 0.04 | 0.07 | 0.03 | 0.01 | 0.10 | 0.13 | 0.11 | 0.11 | 0.10 | 0.08 | 0.12 | 0.11 | 0.11 |
| H I | 4102 | 0.17 | 0.14 | 0.18 | 0.18 | 0.17 | 0.20 | 0.21 | 0.20 | 0.20 | 0.19 | 0.20 | 0.19 | 0.22 | 0.21 | 0.21 |
| [Fe V]? | 4229 | 0.04 | 0.10 | 0.09 | 0.05 | 0.02 | 0.04 | 0.04 | 0.04 | 0.04 | 0.04 | 0.05 | 0.06 | 0.03 | 0.01 | 0.01 |
| H I | 4340 | 0.33 | 0.32 | 0.35 | 0.42 | 0.40 | 0.44 | 0.42 | 0.44 | 0.44 | 0.38 | 0.39 | 0.41 | 0.42 | 0.42 | 0.43 |
| [O III] | 4363 | <0.07 | <0.04 | <0.04 | 0.05 | 0.04 | 0.08 | 0.07 | 0.07 | 0.07 | 0.10 | 0.11 | 0.13 | 0.11 | 0.12 | 0.12 |
| He II | 4686 | <0.05 | <0.01 | <0.03 | 0.05 | 0.04 | 0.03 | 0.09 | 0.09 | 0.09 | 0.14 | 0.15 | 0.14 | 0.23 | 0.23 | 0.24 |
| [Ar IV] | 4711 | <0.02 | <0.02 | <0.04 | 0.03 | <0.02 | <0.03 | 0.02 | 0.02 | 0.01 | 0.04 | 0.04 | 0.05 | 0.06 | 0.06 | 0.06 |
| H I | 4861 | 1.00 | 1.00 | 1.00 | 1.00 | 1.00 | 1.00 | 1.00 | 1.00 | 1.00 | 1.00 | 1.00 | 1.00 | 1.00 | 1.00 | 1.00 |
| [O III] | 4959 | 0.17 | 0.18 | 0.14 | 0.60 | 0.60 | 0.50 | 1.31 | 1.23 | 1.11 | 1.85 | 1.81 | 1.71 | 2.76 | 2.86 | 2.92 |
| [O III] | 5007 | 0.53 | 0.40 | 0.39 | 1.52 | 1.60 | 1.39 | 2.94 | 3.25 | 2.88 | 5.11 | 4.76 | 4.62 | 8.34 | 8.64 | 8.76 |
| [N I] | 5200 | 0.05 | 0.07 | 0.08 | 0.04 | 0.08 | 0.08 | 0.06 | 0.08 | 0.07 | 0.06 | 0.07 | 0.07 | 0.07 | 0.07 | 0.07 |
| [N II] | 5755 | <0.07 | <0.07 | <0.04 | <0.06 | <0.07 | <0.06 | 0.03 | 0.05 | 0.04 | 0.03 | 0.03 | 0.04 | 0.02 | 0.02 | 0.02 |
| He I | 5876 | 0.14 | 0.08 | 0.07 | 0.11 | 0.09 | 0.09 | 0.12 | 0.11 | 0.11 | 0.13 | 0.12 | 0.11 | 0.12 | 0.12 | 0.12 |
| Fe VII | 6087 | <0.00 | <0.01 | <0.00 | <0.01 | <0.01 | <0.01 | 0.02 | 0.03 | 0.03 | 0.04 | 0.05 | 0.07 | 0.08 | 0.08 | 0.08 |
| [O I] | 6300 | 0.24 | 0.25 | 0.16 | 0.23 | 0.23 | 0.22 | 0.33 | 0.35 | 0.31 | 0.35 | 0.31 | 0.33 | 0.40 | 0.40 | 0.40 |
| [O I] | 6363 | 0.10 | 0.06 | 0.04 | 0.07 | 0.06 | 0.06 | 0.10 | 0.10 | 0.10 | 0.11 | 0.13 | 0.14 | 0.11 | 0.11 | 0.11 |
| [N II] | 6548 | 1.05 | 1.35 | 1.30 | 0.91 | 1.15 | 1.40 | 0.97 | 1.20 | 1.35 | 1.03 | 1.19 | 1.34 | 0.94 | 0.95 | 0.98 |
| H I | 6563 | 6.58 | 6.79 | 5.82 | 4.49 | 4.73 | 5.11 | 4.35 | 4.45 | 4.58 | 3.92 | 4.03 | 4.29 | 3.51 | 3.52 | 3.54 |
| [N II] | 6584 | 3.36 | 4.13 | 3.78 | 2.85 | 3.34 | 3.92 | 2.92 | 3.44 | 3.55 | 2.88 | 3.22 | 3.45 | 2.58 | 2.54 | 2.64 |
| He I | 6678 | 0.04 | 0.03 | 0.02 | 0.02 | 0.02 | 0.04 | 0.04 | 0.03 | 0.03 | 0.04 | 0.04 | 0.03 | 0.04 | 0.04 | 0.04 |
| [S II] | 6716 | 1.25 | 1.21 | 0.89 | 0.90 | 0.86 | 0.86 | 0.93 | 0.93 | 0.88 | 0.97 | 0.94 | 0.92 | 1.01 | 0.96 | 0.96 |
| [S II] | 6731 | 0.98 | 1.01 | 0.76 | 0.76 | 0.76 | 0.75 | 0.80 | 0.83 | 0.78 | 0.85 | 0.84 | 0.82 | 0.85 | 0.81 | 0.81 |
| [Ar III] | 7135 | 0.08 | 0.10 | 0.06 | 0.09 | 0.09 | 0.09 | 0.14 | 0.14 | 0.14 | 0.19 | 0.18 | 0.18 | 0.25 | 0.25 | 0.25 |
| [O II] | 7325 | 0.13 | 0.12 | 0.07 | 0.08 | 0.13 | 0.17 | 0.16 | 0.19 | 0.13 | 0.16 | 0.15 | 0.16 | 0.15 | 0.16 | 0.16 |

| Table 2b. Dereddened emission line strengths for the AGN locus, relative to Hβ. | | | | | | | | | | | | | | | | |
|---|---|---|---|---|---|---|---|---|---|---|---|---|---|---|---|---|
| Ion | λ | a00 | a01 | a02 | a10 | a11 | a12 | a20 | a21 | a22 | a30 | a31 | a32 | a40 | a41 | a42 |
| [O II] | 3727 | 3.60 | 3.62 | 2.54 | 2.40 | 2.24 | 2.45 | 2.79 | 2.54 | 2.46 | 2.88 | 2.63 | 2.64 | 3.10 | 2.97 | 2.96 |
| H I | 3798 | <0.07 | <0.09 | <0.07 | 0.01 | 0.05 | 0.07 | 0.02 | 0.05 | 0.08 | 0.05 | 0.02 | 0.02 | 0.06 | 0.06 | 0.06 |
| H I | 3835 | <0.03 | <0.04 | <0.10 | 0.03 | 0.04 | 0.06 | 0.01 | 0.05 | 0.06 | 0.05 | 0.02 | 0.04 | 0.09 | 0.09 | 0.10 |
| [Ne III] | 3869 | 0.22 | 0.04 | 0.17 | 0.25 | 0.28 | 0.31 | 0.62 | 0.64 | 0.60 | 0.77 | 0.76 | 0.83 | 0.96 | 0.97 | 1.00 |
| H I | 3889 | 0.08 | 0.07 | 0.15 | 0.17 | 0.15 | 0.19 | 0.22 | 0.22 | 0.22 | 0.17 | 0.18 | 0.21 | 0.20 | 0.19 | 0.20 |
| [S II] | 4070 | 0.23 | 0.12 | 0.07 | 0.10 | 0.05 | 0.02 | 0.14 | 0.18 | 0.16 | 0.14 | 0.13 | 0.11 | 0.14 | 0.13 | 0.13 |
| H I | 4102 | 0.32 | 0.27 | 0.31 | 0.25 | 0.25 | 0.31 | 0.29 | 0.28 | 0.29 | 0.24 | 0.25 | 0.26 | 0.25 | 0.25 | 0.25 |
| [Fe V]? | 4229 | 0.06 | 0.18 | 0.15 | 0.06 | 0.03 | 0.06 | 0.05 | 0.05 | 0.06 | 0.04 | 0.07 | 0.07 | 0.03 | 0.02 | 0.01 |
| H I | 4340 | 0.50 | 0.51 | 0.51 | 0.53 | 0.52 | 0.59 | 0.53 | 0.56 | 0.56 | 0.45 | 0.46 | 0.51 | 0.47 | 0.47 | 0.48 |
| [O III] | 4363 | <0.10 | <0.06 | <0.06 | 0.06 | 0.05 | 0.10 | 0.09 | 0.09 | 0.09 | 0.12 | 0.13 | 0.16 | 0.12 | 0.13 | 0.13 |
| He II | 4686 | <0.06 | <0.01 | <0.03 | 0.05 | 0.04 | 0.03 | 0.10 | 0.10 | 0.10 | 0.15 | 0.16 | 0.15 | 0.24 | 0.24 | 0.24 |
| [Ar IV] | 4711 | <0.02 | <0.02 | <0.05 | 0.04 | <0.03 | <0.04 | 0.03 | 0.02 | 0.01 | 0.04 | 0.04 | 0.05 | 0.06 | 0.06 | 0.06 |
| H I | 4861 | 1.00 | 1.00 | 1.00 | 1.00 | 1.00 | 1.00 | 1.00 | 1.00 | 1.00 | 1.00 | 1.00 | 1.00 | 1.00 | 1.00 | 1.00 |
| [O III] | 4959 | 0.16 | 0.16 | 0.13 | 0.58 | 0.57 | 0.47 | 1.26 | 1.18 | 1.07 | 1.80 | 1.76 | 1.65 | 2.71 | 2.81 | 2.87 |
| [O III] | 5007 | 0.48 | 0.36 | 0.35 | 1.43 | 1.50 | 1.29 | 2.78 | 3.07 | 2.71 | 4.91 | 4.56 | 4.39 | 8.12 | 8.41 | 8.53 |
| [N I] | 5200 | 0.04 | 0.05 | 0.06 | 0.04 | 0.07 | 0.07 | 0.06 | 0.07 | 0.06 | 0.05 | 0.07 | 0.06 | 0.06 | 0.06 | 0.07 |
| [N II] | 5755 | <0.04 | <0.04 | <0.02 | <0.04 | <0.05 | 0.04 | 0.03 | 0.04 | 0.03 | 0.03 | 0.03 | 0.03 | 0.02 | 0.02 | 0.02 |
| He I | 5876 | 0.08 | 0.04 | 0.04 | 0.08 | 0.06 | 0.06 | 0.09 | 0.08 | 0.08 | 0.10 | 0.10 | 0.09 | 0.11 | 0.10 | 0.10 |
| Fe VII | 6087 | <0.01 | 0.01 | <0.01 | 0.01 | <0.01 | <0.01 | 0.02 | 0.02 | 0.02 | 0.03 | 0.05 | 0.07 | 0.07 | 0.07 | 0.08 |
| [O I] | 6300 | 0.11 | 0.12 | 0.09 | 0.15 | 0.15 | 0.13 | 0.23 | 0.23 | 0.21 | 0.26 | 0.23 | 0.23 | 0.33 | 0.34 | 0.33 |
| [O I] | 6363 | 0.05 | 0.03 | 0.02 | 0.05 | 0.04 | 0.03 | 0.07 | 0.07 | 0.07 | 0.08 | 0.09 | 0.10 | 0.09 | 0.09 | 0.09 |
| [N II] | 6548 | 0.46 | 0.57 | 0.64 | 0.58 | 0.70 | 0.79 | 0.64 | 0.77 | 0.85 | 0.75 | 0.85 | 0.90 | 0.77 | 0.77 | 0.79 |
| H I | 6563 | 2.86 | 2.86 | 2.86 | 2.86 | 2.86 | 2.86 | 2.86 | 2.86 | 2.86 | 2.86 | 2.86 | 2.86 | 2.86 | 2.86 | 2.86 |
| [N II] | 6584 | 0.11 | 1.72 | 1.84 | 1.81 | 2.01 | 2.18 | 1.91 | 2.20 | 2.21 | 2.10 | 2.28 | 2.29 | 2.09 | 2.05 | 2.13 |
| He I | 6678 | 0.05 | 0.01 | 0.01 | 0.01 | 0.01 | 0.02 | 0.02 | 0.02 | 0.02 | 0.03 | 0.03 | 0.02 | 0.03 | 0.03 | 0.03 |
| [S II] | 6716 | 0.46 | 0.48 | 0.41 | 0.56 | 0.50 | 0.46 | 0.60 | 0.58 | 0.53 | 0.69 | 0.65 | 0.59 | 0.81 | 0.76 | 0.77 |
| [S II] | 6731 | 2.86 | 0.40 | 0.35 | 0.46 | 0.44 | 0.40 | 0.51 | 0.52 | 0.47 | 0.60 | 0.58 | 0.53 | 0.68 | 0.64 | 0.65 |
| [Ar III] | 7135 | 0.03 | 0.03 | 0.02 | 0.05 | 0.05 | 0.05 | 0.08 | 0.08 | 0.08 | 0.13 | 0.12 | 0.11 | 0.19 | 0.19 | 0.19 |
| [O II] | 7325 | 0.04 | 0.04 | 0.03 | 0.05 | 0.06 | 0.08 | 0.09 | 0.10 | 0.07 | 0.10 | 0.10 | 0.09 | 0.11 | 0.11 | 0.11 |

| Table 3. Emission line ratio predictions – Dust-free Model | | | | | | | | | | |
|---|---|---|---|---|---|---|---|---|---|---|
| | Free Parameters | | | | | Free Parameters | | | | |
| Density Weighting $\beta$ | -1.4 | -1.4 | -1.4 | -1.4 | -1.4 | -1.4 | -1.4 | -1.4 | -1.4 | -1.4 |
| Radial Weighting $\gamma$ | 1.0 | 0.0 | -0.25 | -0.5 | -0.75 | 1.0 | 0.0 | -0.25 | -0.5 | -0.75 |
| | Predicted Overall Properties | | | | | | | | | |
| $W_\lambda(H\beta)$ (Å) | 186 | 194 | 198 | 203 | 209 | | | | | |
| $W_\lambda([O\ III]\ 5007)$ (Å) | 107 | 368 | 539 | 804 | 1189 | | | | | |
| $L_{ion}\Omega$ ($10^{43}$ erg s$^{-1}$) | 6.62 | 2.00 | 2.12 | 1.23 | 0.91 | | | | | |
| $L_{ion}\Omega / L^c([O\ III])$ | 0.47 | 0.11 | 0.05 | 0.04 | 0.02 | | | | | |
| $L_{ion}\Omega / L^c(H\beta)$ | 0.17 | 0.17 | 0.16 | 0.16 | 0.15 | | | | | |
| $r_{max}\Omega^{1/2}$ ($10^{22}$ cm) | 1.44 | 0.80 | 0.82 | 0.62 | 0.54 | | | | | |
| $r_{char}(H\beta)\Omega^{1/2}$ ($10^{22}$ cm) | 1.33 | 0.58 | 0.52 | 0.32 | 0.21 | | | | | |
| $r_{char}([O\ III]\ 5007)\Omega^{1/2}$ ($10^{22}$ cm) | 0.61 | 0.18 | 0.14 | 0.08 | 0.05 | | | | | |
| | $\log_{10}$ ( Predicted Line Ratio) | | | | | $\log_{10}$(Predicted / Observed) Line Ratio | | | | |
| Line ratio | a01 | a11 | a21 | a31 | a41 | a01 | a11 | a21 | a31 | a41 |
| [O II] 3727 / [O III] 5007 | 0.45 | -0.05 | -0.22 | -0.41 | -0.63 | -0.55 | -0.23 | -0.14 | -0.17 | -0.18 |
| [O II] 3727 / [O I] 6300 | 0.70 | 0.73 | 0.74 | 0.75 | 0.74 | -0.80 | -0.46 | -0.29 | -0.31 | -0.20 |
| [O II] 3727 / [O II] 7325 | 2.10 | 1.90 | 1.81 | 1.69 | 1.53 | 0.10 | 0.36 | 0.41 | 0.27 | 0.13 |
| [Ne III] 3869 / He II 4686 | 0.24 | 0.37 | 0.43 | 0.50 | 0.57 | <-0.20 | -0.47 | -0.38 | -0.19 | -0.04 |
| [S II] 4070 / [S II] 6720 | -1.38 | -1.27 | -1.21 | -1.13 | -1.02 | -0.53 | -0.002 | -0.42 | -0.16 | 0.02 |
| [O III] 4363 / [O III] 5007 | -2.52 | -2.38 | -2.34 | -2.29 | -2.22 | >-1.74 | -0.87 | -0.78 | -0.74 | -0.41 |
| He II 4686 / H$\beta$ | -0.90 | -0.84 | -0.81 | -0.77 | -0.74 | >0.92 | 0.55 | 0.20 | 0.03 | -0.11 |
| He II 4686 / He I 5876 | -0.18 | -0.10 | -0.05 | -0.002 | 0.06 | >0.28 | 0.10 | -0.14 | -0.21 | -0.30 |
| [Ar IV] 4711 / [Ar III] 7135 | -2.15 | -1.41 | -1.20 | -0.98 | -0.76 | >-2.00 | -1.11 | -0.59 | -0.49 | -0.25 |
| [O III] 5007 / H$\beta$ | -0.24 | 0.28 | 0.43 | 0.60 | 0.75 | 0.20 | 0.10 | -0.05 | -0.06 | -0.17 |
| [O III] 5007 / [Ar III] 7135 | 0.69 | 1.16 | 1.30 | 1.46 | 1.64 | -0.33 | -0.32 | -0.27 | -0.12 | 0.00 |
| [N I] 5200 / H$\alpha$ | -1.68 | -1.75 | -1.80 | -1.87 | -2.00 | 0.04 | -0.12 | -0.19 | -0.24 | -0.33 |
| [N I] 5200 / [N II] 6584 | -1.43 | -1.48 | -1.51 | -1.54 | -1.59 | 0.07 | -0.01 | -0.02 | -0.01 | -0.07 |
| [N I] 5200 / [O II] 7325 | 0.69 | 0.40 | 0.27 | 0.10 | -0.11 | 0.51 | 0.39 | 0.43 | 0.29 | 0.16 |
| [N II] 5755 / [N II] 6584 | -2.38 | -2.28 | -2.23 | -2.16 | -2.05 | >-0.75 | >-0.68 | -0.50 | -0.20 | -0.02 |
| He I 5876 / H$\beta$ | -0.72 | -0.74 | -0.75 | -0.77 | -0.79 | 0.64 | 0.45 | 0.33 | 0.25 | 0.19 |
| [O I] 6300 / [O III] 5007 | -0.25 | -0.79 | -0.96 | -1.16 | -1.37 | 0.25 | 0.23 | 0.16 | 0.14 | 0.02 |
| [O I] 6300 / H$\alpha$ | -0.97 | -0.99 | -1.00 | -1.03 | -1.09 | 0.42 | 0.31 | 0.08 | 0.06 | -0.16 |
| [O I] 6300 / [N II] 6584 | -0.72 | -0.72 | -0.71 | -0.70 | -0.68 | 0.45 | 0.42 | 0.26 | 0.29 | 0.10 |
| H$\alpha$ / H$\beta$ | 0.48 | 0.48 | 0.48 | 0.48 | 0.47 | 0.03 | 0.02 | 0.02 | 0.02 | 0.02 |
| [N II] 6584 / [O II] 3727 | 0.02 | -0.01 | -0.03 | -0.04 | -0.06 | 0.35 | 0.03 | 0.04 | 0.02 | 0.01 |
| [N II] 6584 / H$\alpha$ | -0.25 | -0.27 | -0.29 | -0.33 | -0.41 | -0.03 | -0.12 | -0.18 | -0.23 | -0.26 |
| [S II] 6716 / [S II] 6731 | 0.08 | 0.06 | 0.05 | 0.04 | 0.03 | -0.01 | 0.00 | 0.00 | -0.01 | -0.05 |
| [S II] 6720 / H$\alpha$ | -0.09 | -0.15 | -0.19 | -0.25 | -0.36 | 0.42 | 0.33 | 0.23 | 0.11 | -0.05 |
| [O II] 7325 / [O I] 6300 | -1.40 | -1.17 | -1.07 | -0.94 | -0.79 | -0.89 | -0.82 | -0.71 | -0.58 | -0.33 |
| [O II] 7325 / H$\alpha$ | -2.37 | -2.16 | -2.07 | -1.98 | -1.88 | -0.47 | -0.51 | -0.62 | -0.52 | -0.49 |
| Fraction of Ratios Fitted | - | - | - | - | - | 0.50 | 0.46 | 0.62 | 0.81 | 0.85 |

| Table 4. Emission line ratio predictions – Dusty Model | | | | | | | | | | |
|---|---|---|---|---|---|---|---|---|---|---|
| | Free Parameters | | | | | Free Parameters | | | | |
| Density Weighting $\beta$ | -1.8 | -1.8 | -1.8 | -1.8 | -1.8 | -1.8 | -1.8 | -1.8 | -1.8 | -1.8 |
| Radial Weighting $\gamma$ | 0.5 | -1.0 | -1.25 | -1.5 | -1.75 | 0.5 | -1.0 | -1.25 | -1.5 | -1.75 |
| | Predicted Equivalent Widths | | | | | | | | | |
| $W_\lambda$(H$\beta$) (Å) | 49.1 | 26.2 | 18.1 | 11.4 | 6.98 | | | | | |
| $W_\lambda$([O III] 5007) (Å) | 32.2 | 97.5 | 89.4 | 73.6 | 56.1 | | | | | |
| $L_{ion}\Omega$ ($10^{43}$ erg s$^{-1}$) | 5.96 | 3.53 | 5.55 | 5.18 | 6.51 | | | | | |
| $L_{ion}\Omega / L^c$([O III]) | 0.43 | 0.20 | 0.14 | 0.15 | 0.13 | | | | | |
| $L_{ion}\Omega / L^c$(H$\beta$) | 0.16 | 0.29 | 0.43 | 0.67 | 1.10 | | | | | |
| $r_{max}\Omega^{1/2}$ ($10^{22}$ cm) | 1.37 | 1.06 | 1.32 | 1.28 | 1.44 | | | | | |
| $r_{char}$(H$\beta$)$\Omega^{1/2}$ ($10^{22}$ cm) | 1.22 | 0.53 | 0.53 | 0.38 | 0.29 | | | | | |
| $r_{char}$([O III] 5007) $\Omega^{1/2}$ ($10^{22}$ cm) | 0.55 | 0.16 | 0.16 | 0.12 | 0.10 | | | | | |
| | Model Predictions (log of line ratio) | | | | | Log(Predicted / Observed) | | | | |
| Line ratio | a01 | a11 | a21 | a31 | a41 | a01 | a11 | a21 | a31 | a41 |
| [O II] 3727 / [O III] 5007 | 0.79 | -0.08 | -0.26 | -0.45 | -0.65 | -0.21 | -0.26 | -0.18 | -0.22 | -0.20 |
| [O II] 3727 / [O I] 6300 | 0.77 | 0.78 | 0.76 | 0.72 | 0.64 | -0.49 | -0.41 | -0.27 | -0.34 | -0.30 |
| [O II] 3727 / [O II] 7325 | 1.94 | 1.76 | 1.68 | 1.57 | 1.41 | -0.06 | 0.23 | 0.28 | 0.15 | -0.002 |
| [Ne III] 3869 / He II 4686 | 0.63 | 0.57 | 0.54 | 0.51 | 0.48 | <0.18 | -0.27 | -0.27 | -0.18 | -0.14 |
| [S II] 4070 / [S II] 6720 | -1.24 | -1.18 | -1.15 | -1.12 | -1.06 | -0.38 | 0.09 | -0.36 | -0.15 | 0.02 |
| [O III] 4363 / [O III] 5007 | -2.32 | -1.73 | -1.60 | -1.46 | -1.31 | >-1.54 | -0.22 | -0.05 | 0.08 | 0.50 |
| He II 4686 / H$\beta$ | -1.37 | -1.07 | -0.98 | -0.87 | -0.77 | >0.45 | 0.32 | 0.03 | -0.07 | -0.14 |
| He II 4686 / He I 5876 | -0.53 | -0.21 | -0.12 | 0.01 | 0.11 | >-0.07 | 0.02 | -0.20 | -0.22 | -0.25 |
| [Ar IV] 4711 / [Ar III] 7135 | -2.35 | -1.25 | -1.09 | -0.95 | -0.84 | >-2.11 | -0.95 | -0.49 | -0.47 | -0.33 |
| [O III] 5007 / H$\beta$ | -0.18 | 0.57 | 0.69 | 0.81 | 0.90 | 0.26 | 0.39 | 0.21 | 0.15 | -0.02 |
| [O III] 5007 / [Ar III] 7135 | 0.47 | 1.19 | 1.32 | 1.43 | 1.55 | -0.54 | -0.28 | -0.26 | -0.14 | -0.10 |
| [N I] 5200 / H$\alpha$ | -1.13 | -1.30 | -1.36 | -1.43 | -1.51 | 0.59 | 0.33 | 0.24 | 0.21 | 0.15 |
| [N I] 5200 / [N II] 6584 | -1.21 | -1.24 | -1.24 | -1.23 | -1.22 | 0.29 | 0.24 | 0.25 | 0.31 | 0.30 |
| [N I] 5200 / [O II] 7325 | 0.68 | 0.46 | 0.38 | 0.28 | 0.15 | 0.50 | 0.45 | 0.54 | 0.47 | 0.43 |
| [N II] 5755 / [N II] 6584 | -2.11 | -1.97 | -1.89 | -1.75 | -1.55 | >-0.49 | >-0.37 | -0.16 | 0.21 | 0.49 |
| He I 5876 / H$\beta$ | -0.84 | -0.85 | -0.86 | -0.87 | -0.88 | 0.52 | 0.34 | 0.23 | 0.15 | 0.10 |
| [O I] 6300 / [O III] 5007 | 0.02 | -0.86 | -1.03 | -1.17 | -1.30 | 0.52 | 0.15 | 0.09 | 0.12 | 0.10 |
| [O I] 6300 / H$\alpha$ | -0.64 | -0.78 | -0.83 | -0.87 | -0.90 | 0.75 | 0.51 | 0.26 | 0.22 | 0.03 |
| [O I] 6300 / [N II] 6584 | -0.73 | -0.72 | -0.70 | -0.67 | -0.62 | 0.45 | 0.42 | 0.27 | 0.32 | 0.17 |
| H$\alpha$ / H$\beta$ | 0.48 | 0.49 | 0.49 | 0.50 | 0.51 | 0.02 | 0.03 | 0.04 | 0.04 | 0.05 |
| [N II] 6584 / [O II] 3727 | 0.05 | -0.06 | -0.06 | -0.05 | -0.03 | 0.27 | -0.02 | 0.003 | 0.01 | 0.13 |
| [N II] 6584 / H$\alpha$ | 0.08 | -0.06 | -0.12 | -0.20 | -0.29 | 0.30 | 0.09 | -0.01 | -0.10 | -0.14 |
| [S II] 6716 / [S II] 6731 | 0.09 | 0.08 | 0.07 | 0.06 | 0.06 | 0.01 | 0.02 | 0.02 | 0.02 | -0.02 |
| [S II] 6720 / H$\alpha$ | 0.01 | -0.20 | -0.28 | -0.38 | -0.49 | 0.52 | 0.28 | 0.14 | <0.001 | -0.18 |
| [O II] 7325 / [O I] 6300 | -1.17 | -0.98 | -0.92 | -0.85 | -0.76 | -0.66 | -0.63 | -0.56 | -0.48 | -0.30 |
| [O II] 7325 / H$\alpha$ | -1.81 | -1.77 | -1.74 | -1.71 | -1.67 | 0.09 | -0.12 | -0.30 | -0.26 | -0.27 |
| Fraction of Ratios Fitted | - | - | - | - | - | 0.54 | 0.62 | 0.85 | 0.77 | 0.85 |

| Table 5. Emission line ratio predictions – Mixing Model | | | | | | | | | | |
|---|---|---|---|---|---|---|---|---|---|---|
| | Free Parameters | | | | | Free Parameters | | | | |
| AGN Fraction | 0.02 | 0.16 | 0.35 | 0.53 | 1.00 | 0.02 | 0.16 | 0.35 | 0.53 | 1.00 |
| | Model Predictions (log of line ratio) | | | | | Log(Predicted / Observed) | | | | |
| Line ratio | a01 | a11 | a21 | a31 | a41 | a01 | a11 | a21 | a31 | a41 |
| [O II] 3727 / [O III] 5007 | 0.65 | 0.08 | -0.17 | -0.29 | -0.45 | -0.35 | -0.09 | -0.09 | -0.05 | 0.00 |
| [O II] 3727 / [O I] 6300 | 1.46 | 1.27 | 1.14 | 1.06 | 0.94 | -0.02 | 0.10 | 0.09 | 0.00 | 0.00 |
| [O II] 3727 / [O II] 7325 | 1.71 | 1.63 | 1.55 | 1.51 | 1.43 | -0.25 | 0.05 | 0.15 | 0.09 | 0.00 |
| [Ne III] 3869 / He II 4686 | >0.90 | >0.71 | >0.65 | >0.63 | >0.61 | <0.30 | <-0.14 | <-0.16 | <-0.05 | <0.00 |
| [S II] 4070 / [S II] 6720 | <-1.03 | <-1.03 | <-1.03 | <-1.03 | <-1.03 | >-0.16 | >0.24 | >-0.24 | >-0.06 | >0.00 |
| [O III] 4363 / [O III] 5007 | <-1.46 | <-1.71 | <-1.77 | <-1.79 | <-1.81 | >-0.68 | >-0.23 | >-0.24 | >-0.25 | >0.00 |
| He II 4686 / Hβ | <-1.82 | <-1.33 | <-1.04 | <-0.88 | <-0.62 | >0.18 | >0.07 | >-0.04 | >-0.08 | >0.00 |
| He II 4686 / He I 5876 | <-0.53 | <-0.09 | <0.13 | <0.24 | <0.38 | >0.07 | >0.08 | >0.03 | >0.03 | >0.00 |
| [Ar IV] 4711 / [Ar III] 7135 | <-0.10 | <-0.33 | <-0.42 | <-0.46 | <-0.50 | >0.08 | >-0.11 | >0.18 | >0.02 | >0.00 |
| [O III] 5007 / Hβ | -0.44 | 0.18 | 0.49 | 0.66 | 0.92 | 0.00 | 0.00 | 0.00 | 0.00 | 0.00 |
| [O III] 5007 / [Ar III] 7135 | 1.41 | 1.59 | 1.62 | 1.63 | 1.65 | 0.33 | 0.11 | 0.04 | 0.05 | 0.00 |
| [N I] 5200 / Hα | -1.85 | -1.82 | -1.78 | -1.75 | -1.68 | -0.09 | -0.21 | -0.17 | -0.14 | 0.00 |
| [N I] 5200 / [N II] 6584 | -1.45 | -1.47 | -1.49 | -1.50 | -1.53 | 0.08 | -0.01 | 0.01 | 0.01 | 0.00 |
| [N I] 5200 / [O II] 7325 | 0.11 | 0.00 | -0.09 | -0.16 | -0.26 | 0.01 | -0.06 | 0.06 | 0.00 | 0.00 |
| [N II] 5755 / [N II] 6584 | <-2.05 | <-2.04 | <-2.03 | <-2.02 | <-2.01 | >-0.42 | >-0.44 | >-0.29 | >-0.14 | >0.00 |
| He I 5876 / Hβ | -1.29 | -1.24 | -1.17 | -1.12 | -1.00 | 0.11 | -0.01 | -0.07 | -0.12 | 0.00 |
| [O I] 6300 / [O III] 5007 | -0.81 | -1.19 | -1.31 | -1.35 | -1.39 | -0.33 | -0.19 | -0.18 | -0.05 | 0.00 |
| [O I] 6300 / Hα | -1.71 | -1.47 | -1.27 | -1.15 | -0.92 | -0.33 | -0.19 | -0.18 | -0.05 | 0.00 |
| [O I] 6300 / [N II] 6584 | -1.31 | -1.12 | -0.98 | -0.90 | -0.78 | -0.15 | 0.01 | 0.00 | 0.10 | 0.00 |
| Hα / Hβ | 0.46 | 0.46 | 0.46 | 0.46 | 0.46 | 0.00 | 0.00 | 0.00 | 0.00 | 0.00 |
| [N II] 6584 / [O II] 3727 | -0.15 | -0.15 | -0.15 | -0.16 | -0.16 | 0.17 | -0.10 | -0.09 | -0.09 | 0.00 |
| [N II] 6584 / Hα | -0.40 | -0.35 | -0.29 | -0.25 | -0.14 | -0.17 | -0.20 | -0.18 | -0.15 | 0.00 |
| [S II] 6716 / [S II] 6731 | 0.13 | 0.12 | 0.10 | 0.09 | 0.07 | 0.05 | 0.06 | 0.06 | 0.04 | 0.00 |
| [S II] 6720 / Hα | -0.64 | -0.57 | -0.50 | -0.44 | -0.31 | -0.13 | -0.09 | -0.08 | -0.07 | 0.00 |
| [O II] 7325 / [O I] 6300 | -0.25 | -0.35 | -0.42 | -0.45 | -0.49 | 0.23 | 0.05 | -0.06 | -0.09 | 0.00 |
| [O II] 7325 / Hα | -1.95 | -1.82 | -1.69 | -1.60 | -1.41 | -0.10 | -0.15 | -0.24 | -0.14 | 0.00 |
| Fraction of Ratios Fitted | - | - | - | - | - | 0.84 | 1.0 | 1.0 | 1.0 | 1.0 |

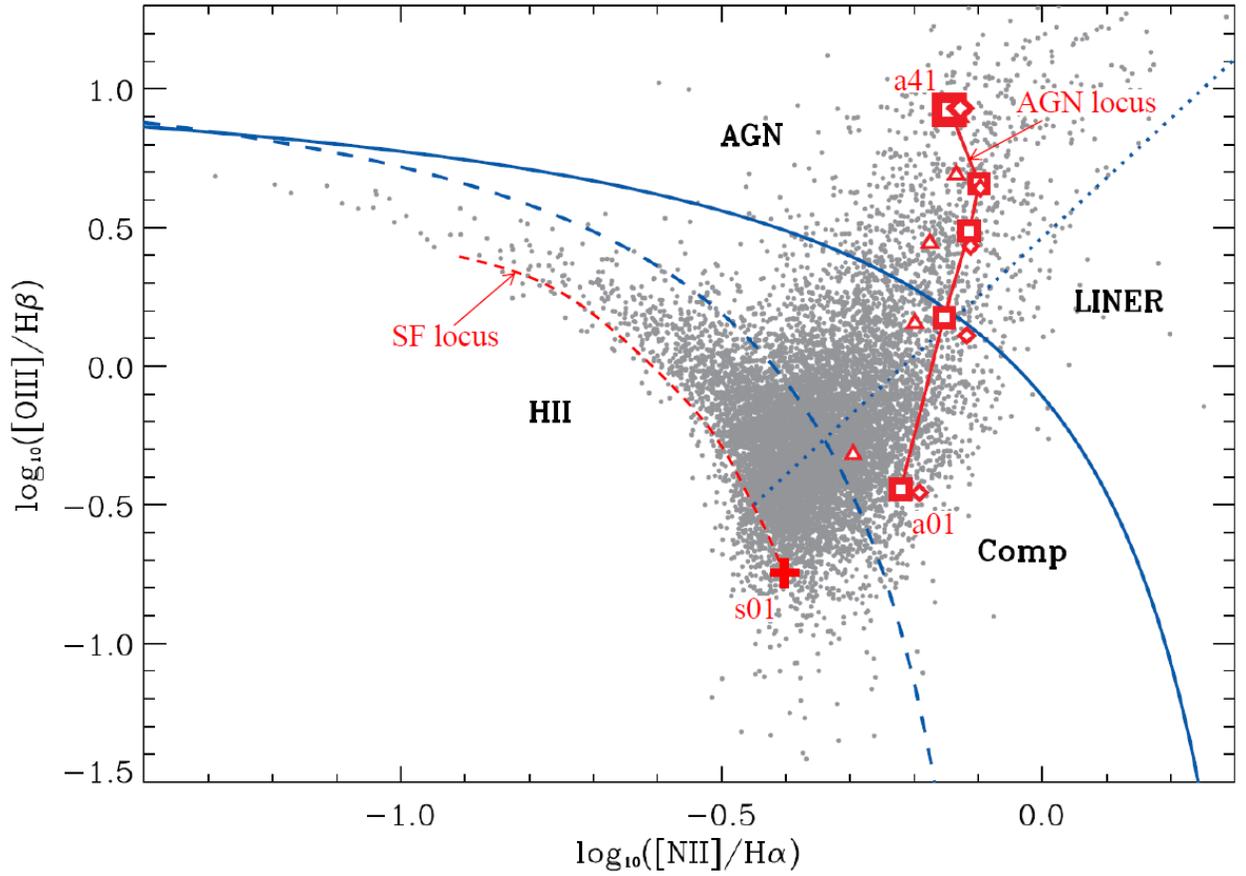

**Figure 1.** The BPT diagram, with our sample of galaxies plotted as grey dots. The solid blue line is the theoretical upper limit for SF galaxies presented by Kewley et al. (2001). The dashed blue line is the classification curve used by Kauffmann et al. (2003) as a lower limit for finding AGN. The dotted blue line shows the division of AGN and LINERs from Kauffmann et al. (2003). The solid red curve is the AGN locus discussed in §1.2, and the dashed red curve is the SF locus. The red symbols show our sequences of AGN picked out by MFICA with a4*j* representing the highest ionization observations, and also the position of the low-ionization SF subsample s01 that is used in §5.4.

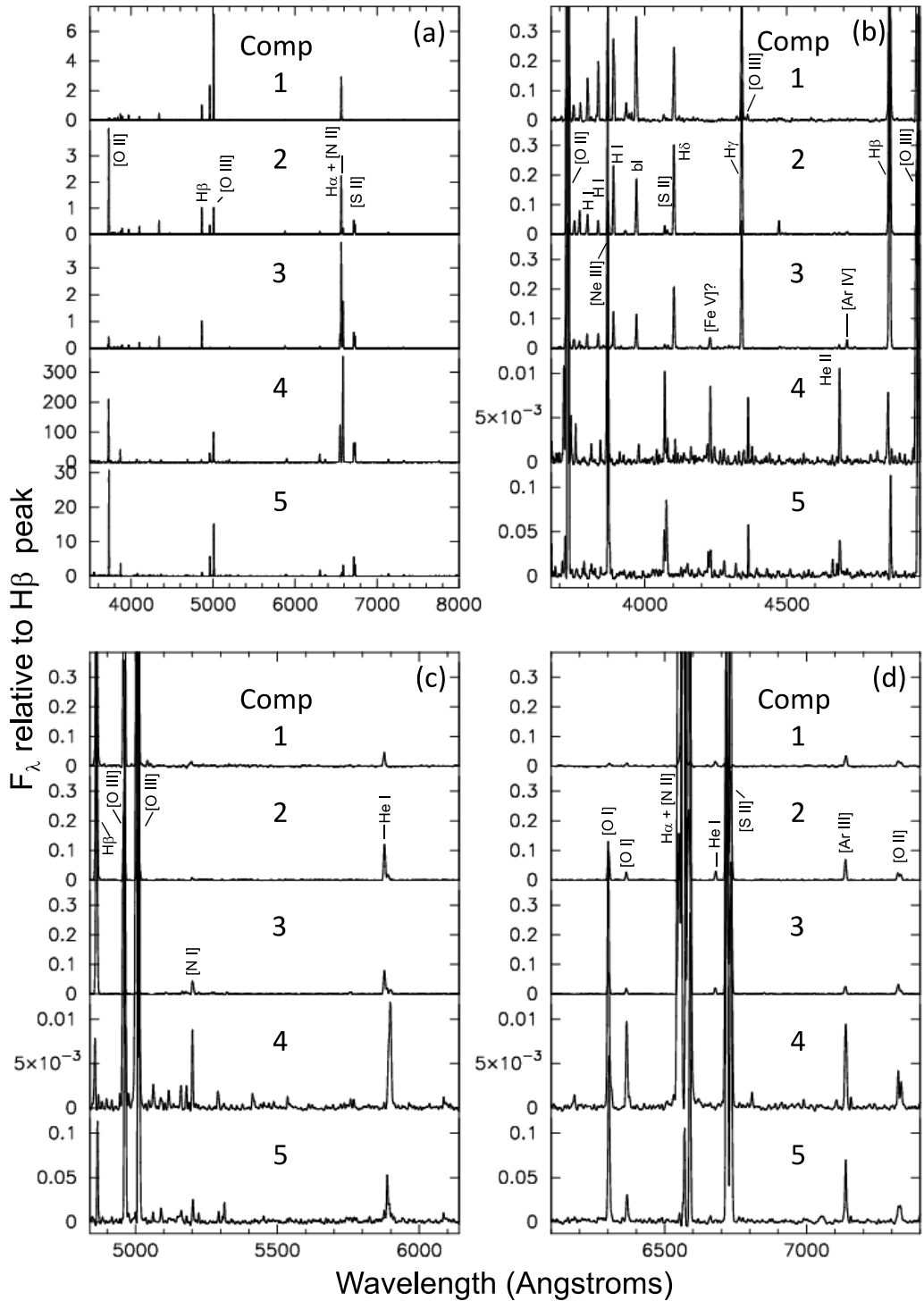

**Figure 2.** Spectra for the five emission line components generated by MFICA. Panel (a) shows the full spectra. The other panels show enlargements that slightly overlap in wavelength. The emission lines listed in Table 2 are identified by their ions. "bl" indicates a blended line. All flux values are shown in units of the peak Hβ intensity in the particular component.

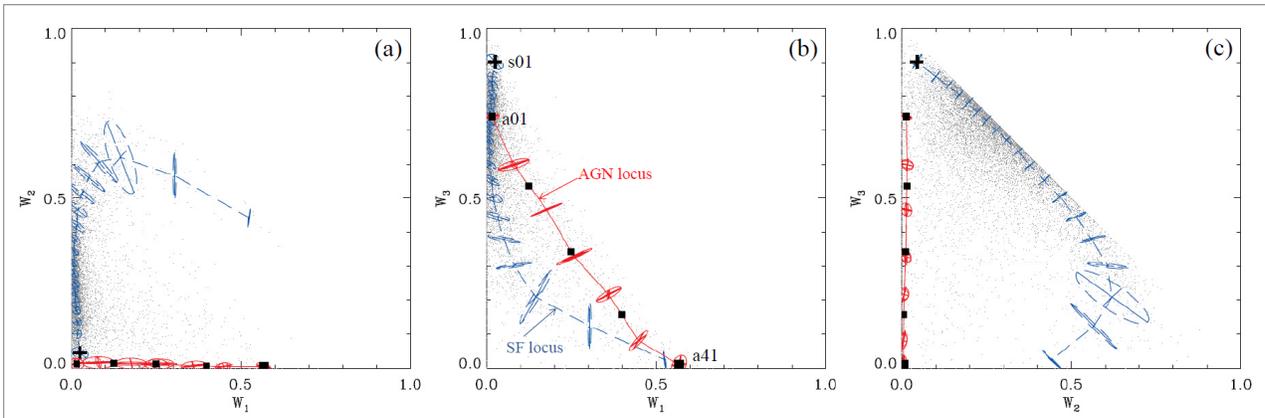

**Figure 3.** Distributions of the MFICA weights for the first three components. In this representation the AGN locus appears as the solid red line, and the SF locus is the dashed blue line. The ovals along each locus represent the scatter of data points still identified as AGN or pure SF, and the sequences described in the text as "wings" (a$i$1 and a$i$2) would follow the outer boundaries of these ovals for the AGN locus. The locations of the AGN subsets described in § 2 are marked by black squares, and the location of the low-ionization SF subset s01 used in § 5.4 is marked by a black cross.

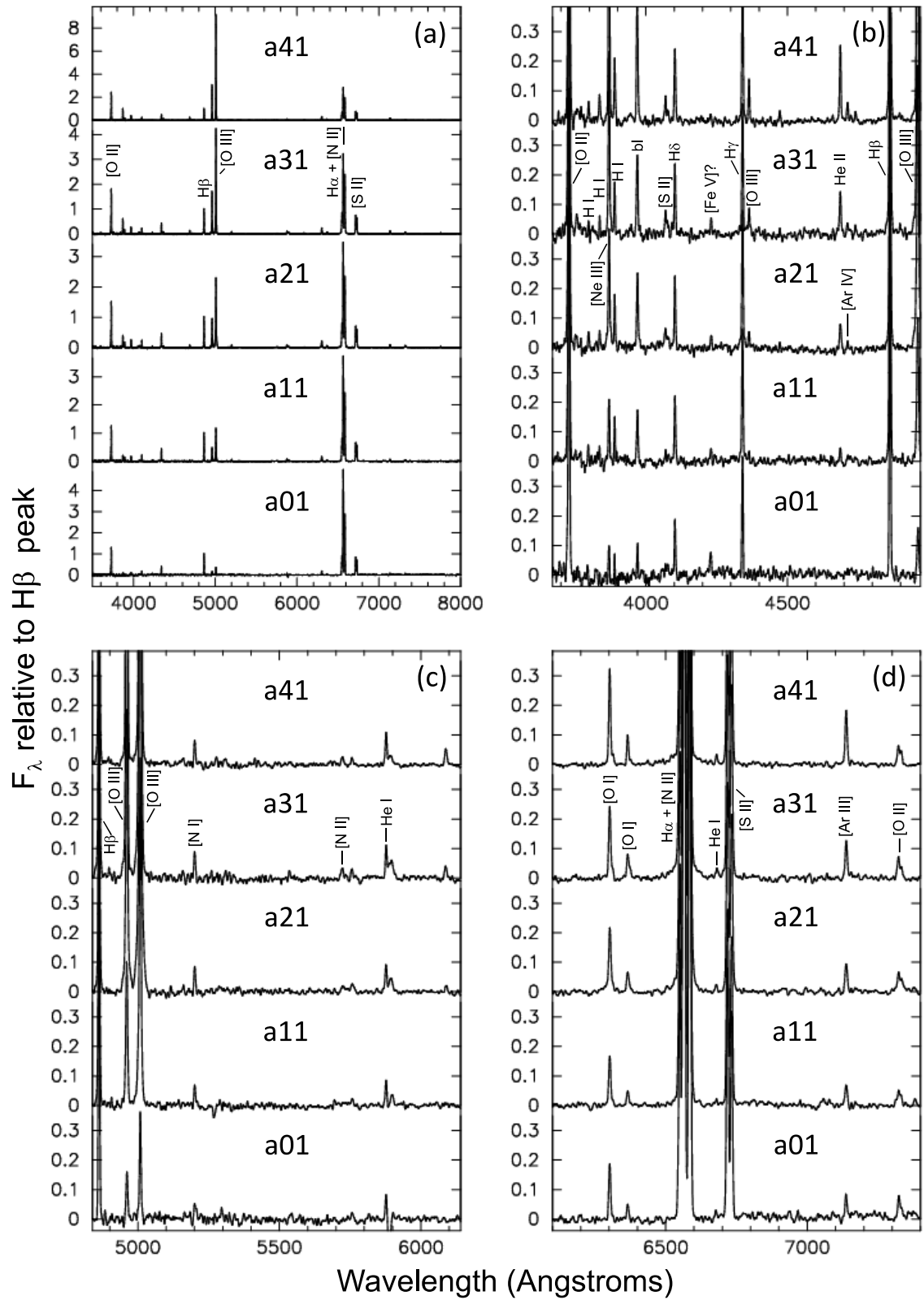

**Figure 4.** Observed, coadded spectra for the five subsets that fall directly along central the AGN locus. Panel (a) shows the full spectra. The other panels show enlargements that slightly overlap in wavelength. The emission lines listed in Table 2 are identified by their ions. "bl" indicates a blended line. All flux values are shown in units of the peak Hβ intensity in the particular spectrum.

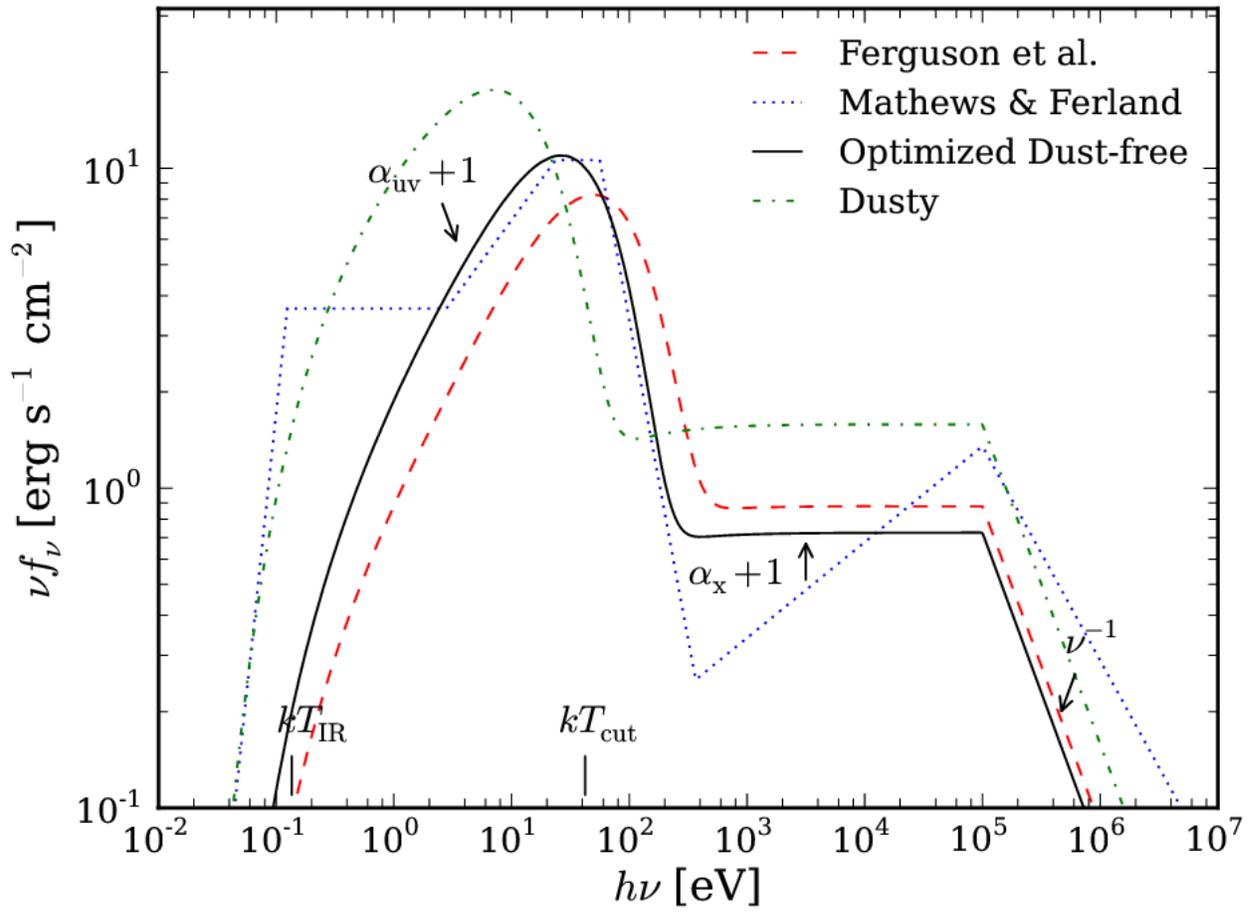

**Figure 5.** The SEDs for Ferguson et al. (1997), Mathews and Ferland (1987), our best model (optimized dust-free) discussed in §3.4.2 and also the SED for our dusty model (§4). Our optimized dust-free models result from an SED that has a shifted UV "Big Bump" in order to match He II λ4686, a well known indicator of the SED. The labeled parts of the SED correspond to the various components for our optimized dust-free model.

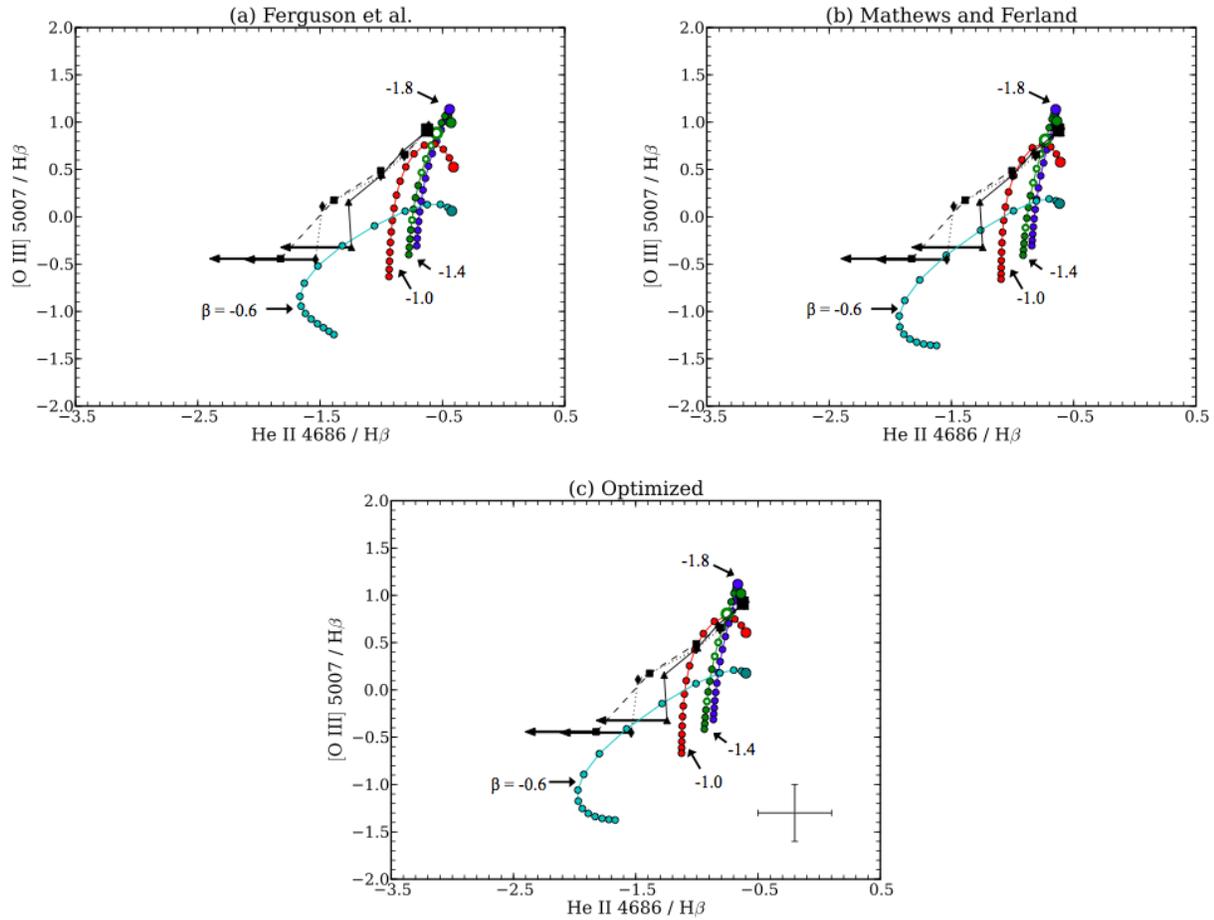

**Figure 6.** The He II λ4686 /Hβ line ratio for various SEDs: Ferguson et al. (top left panel), Mathews and Ferland (top right panel), and our optimized SED (lower panel). In each panel the triangles connected by the solid black line are the dereddened measurements from the co-added spectra that constitute the a40-a30-a20-a10-a00 sequence. Similarly, the squares connected by the dashed black line show the a41-a31-a21-a11-a01 sequence, and the diamonds connected by the dotted black line show the a42-a32-a22-a12-a02 sequence. The largest shape in each sequence indicates observations at the top of AGN locus (a40, a41, a42). The colored lines with circles represent LOC integrations. The density-weighting indices in the LOC integrations, β, are indicated by different colors. The radial weighting indices, γ, become more positive as a function of increasing distance from the largest solid colored circle, for each particular density weighting, covering the range $-2.0 \leq \gamma \leq 2.0$. The hollow circles indicate our best fitting set of free parameters, with the largest hollow circle indicating our best fit to the a41 subset (large black square). Our best fitting free parameters starting from low ionization are γ =1.0, 0.0, -0.25, -0.5, and -0.75 with β = -1.4. The cross in the corner of the plot located in the lower right corner of panel (c) indicates the factor of two range of acceptable error. The He II λ4686 /Hβ line ratio successfully matches the higher to moderate ionization AGN subsets within a factor of two for our optimized SED and the Mathews and Ferland SED.

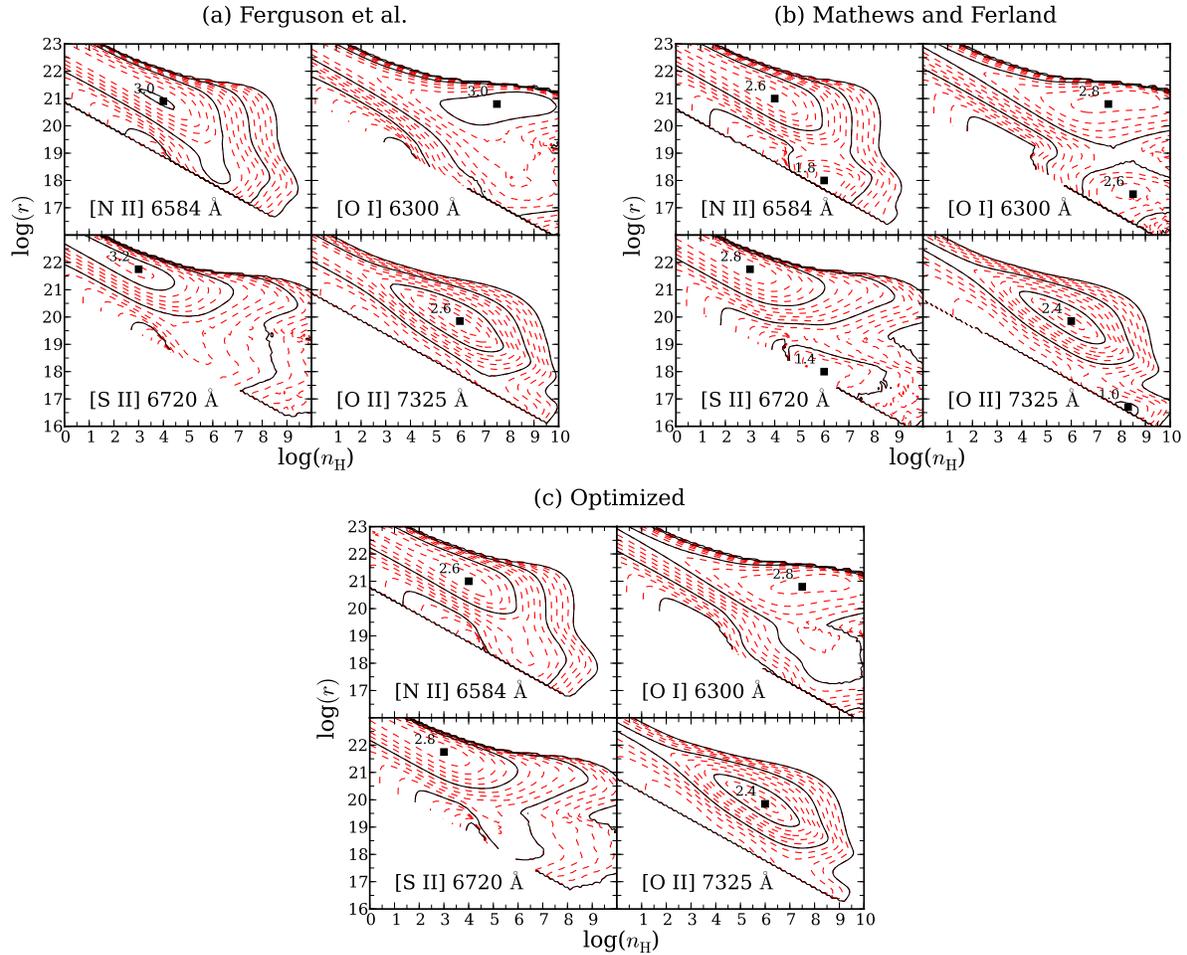

**Figure 7.** Contours for log equivalent widths as a function of radius, $r$, and hydrogen density, $n_H$. All equivalent widths are relative to the continuum at 4860 Å. The contour increments represent steps of 0.2 dex with a lower cutoff at 1 Å. The squares are labeled with log equivalent width and indicate the local maxima for each grid. We show equivalent width diagrams for three different SEDs: (a) Ferguson et al. (1997), (b) Mathews and Ferland (1987), (c) our optimized SED. Panel (b) clearly shows a second region in which the gas is optimally reprocessed into emission lines. This can be seen to a lesser degree in panels (a) and (c) where a "shelf" starts to form in high density, high flux regions.

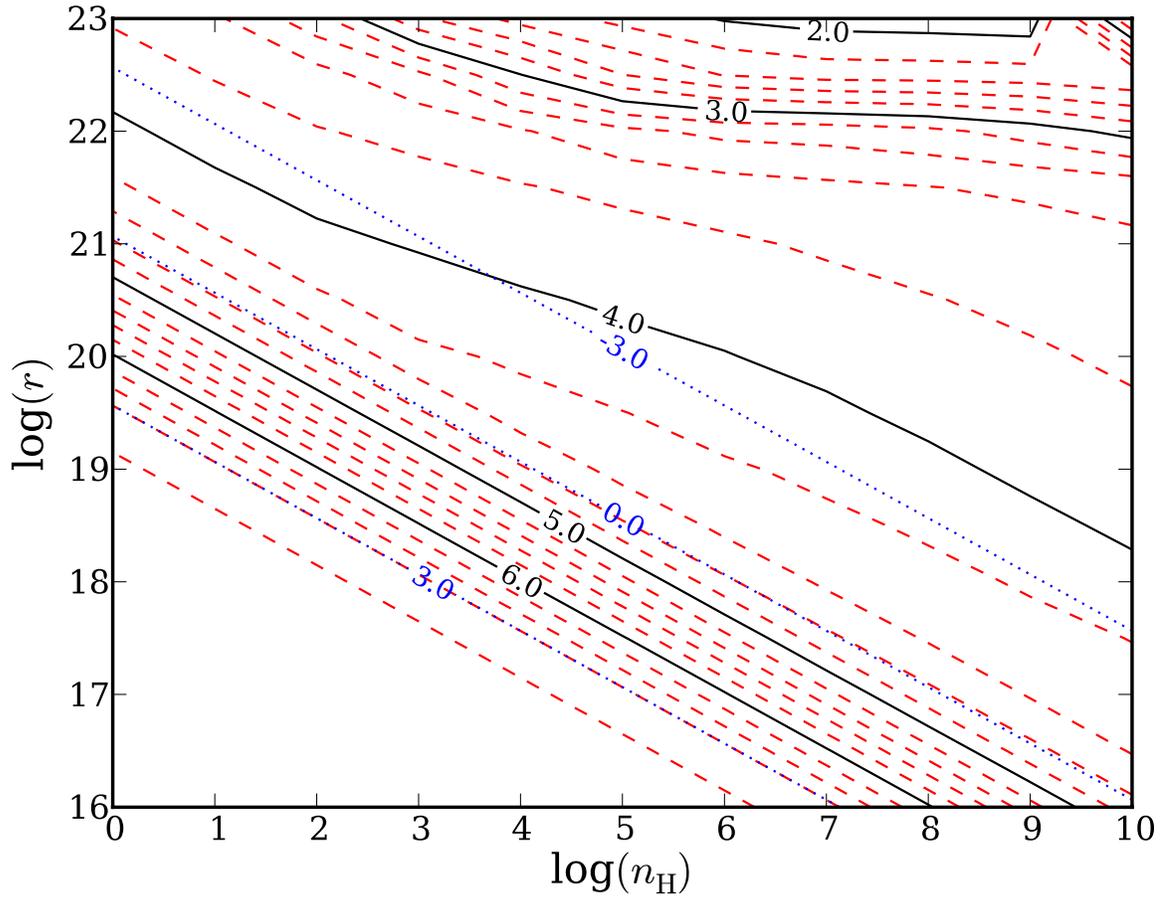

**Figure 8.** The ionization parameter and temperature at the illuminated face of each cloud. The ionization parameter is shown by a representative sample of evenly spaced blue, dotted contours across the LOC plane which are labeled in blue. The temperature is given in 1.0 dex (black lines and labels) and 0.2 dex (red lines) increments.

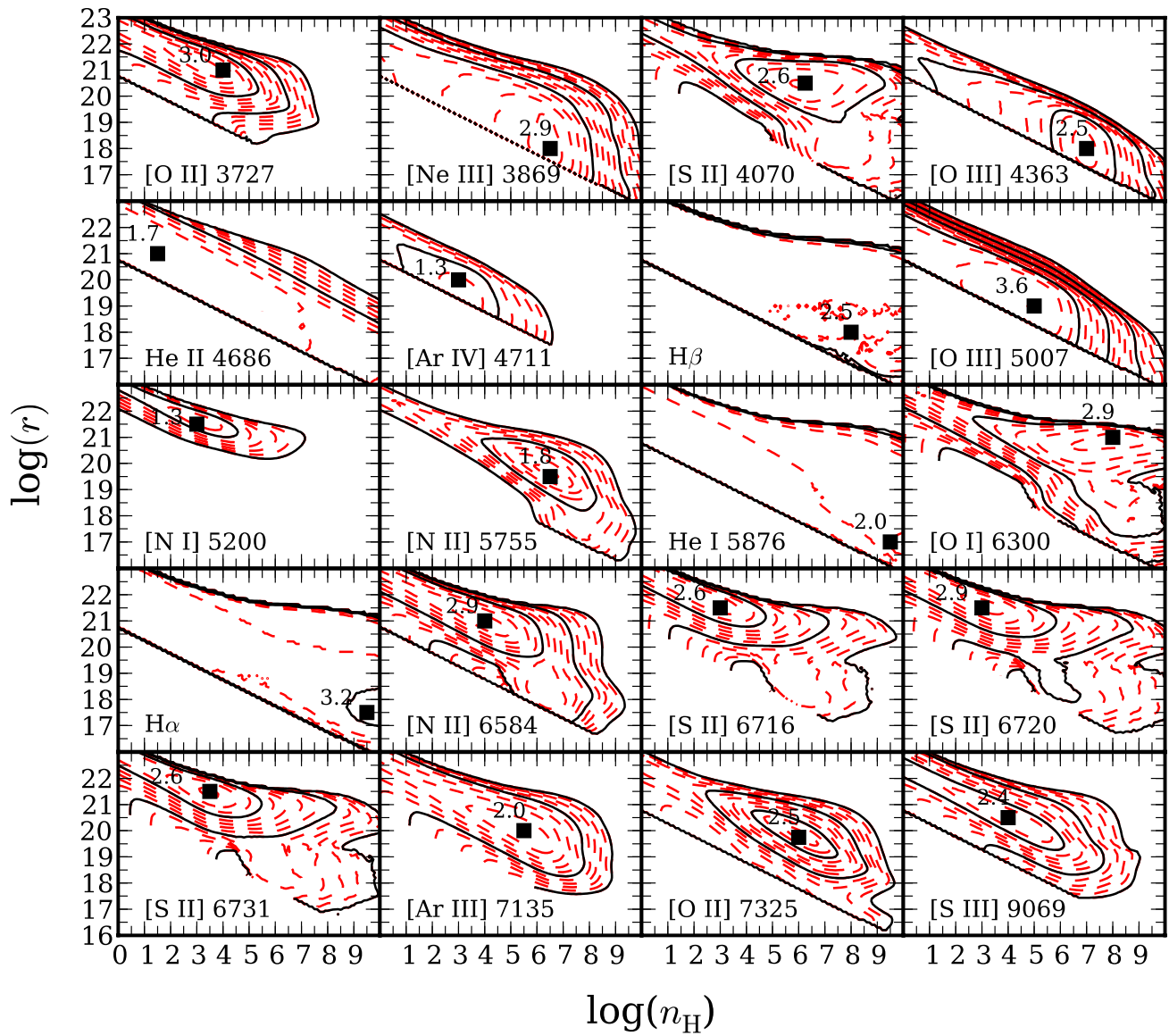

**Figure 9.** Contours of log equivalent width as a function of radius, $r$, and hydrogen density, $n_H$, in the same manner as Fig. 7 but for our best fitting model.

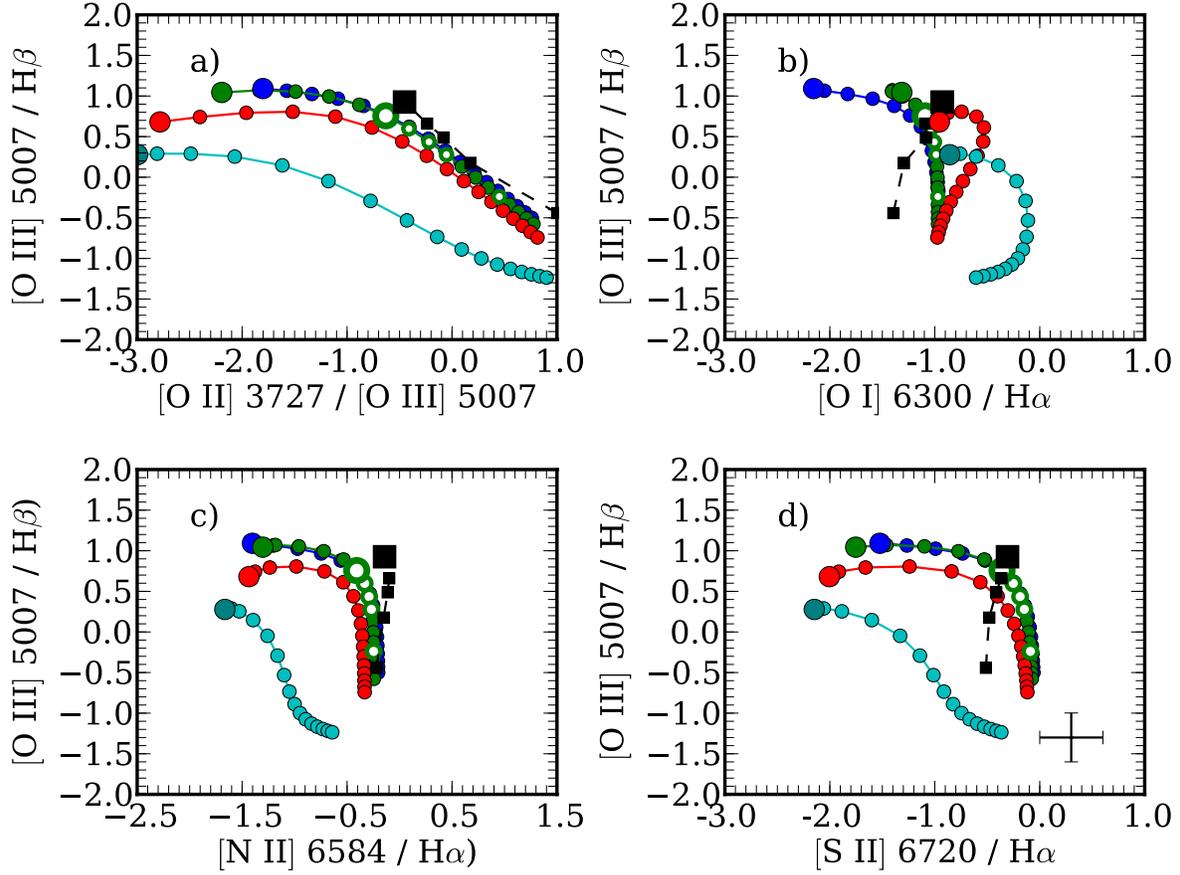

**Figure 10.** Line ratio diagrams constraining the excitation mechanism. In this and all subsequent similar figures, the lines of different colors signify different values of β (the density power law index) with cyan: β = -0.6, red: β = -1.0, green: β = -1.4 and blue: β = -1.8. Along each line, the colored circles represent individual LOC models with different values of γ (the radial distribution power law index) in steps of -0.25, with the large circle at one end of each line always corresponding to the smallest value, γ = -2.0. The black squares are the measured values along the AGN locus, and the open circles indicate the best-fitting LOC models described in Sect. 3.4, with the largest square and largest open circle indicating the highest-ionization point along the AGN sequence. As discussed in the text, the error bars in panel (d) indicate a factor of two uncertainty. Our models successfully fit the high to moderation ionization AGN subsets. In the cases of [N II] λ6584/ Hα and [O III] λ5007/ Hβ we have successfully matched the entire AGN sequence to within a factor of two.

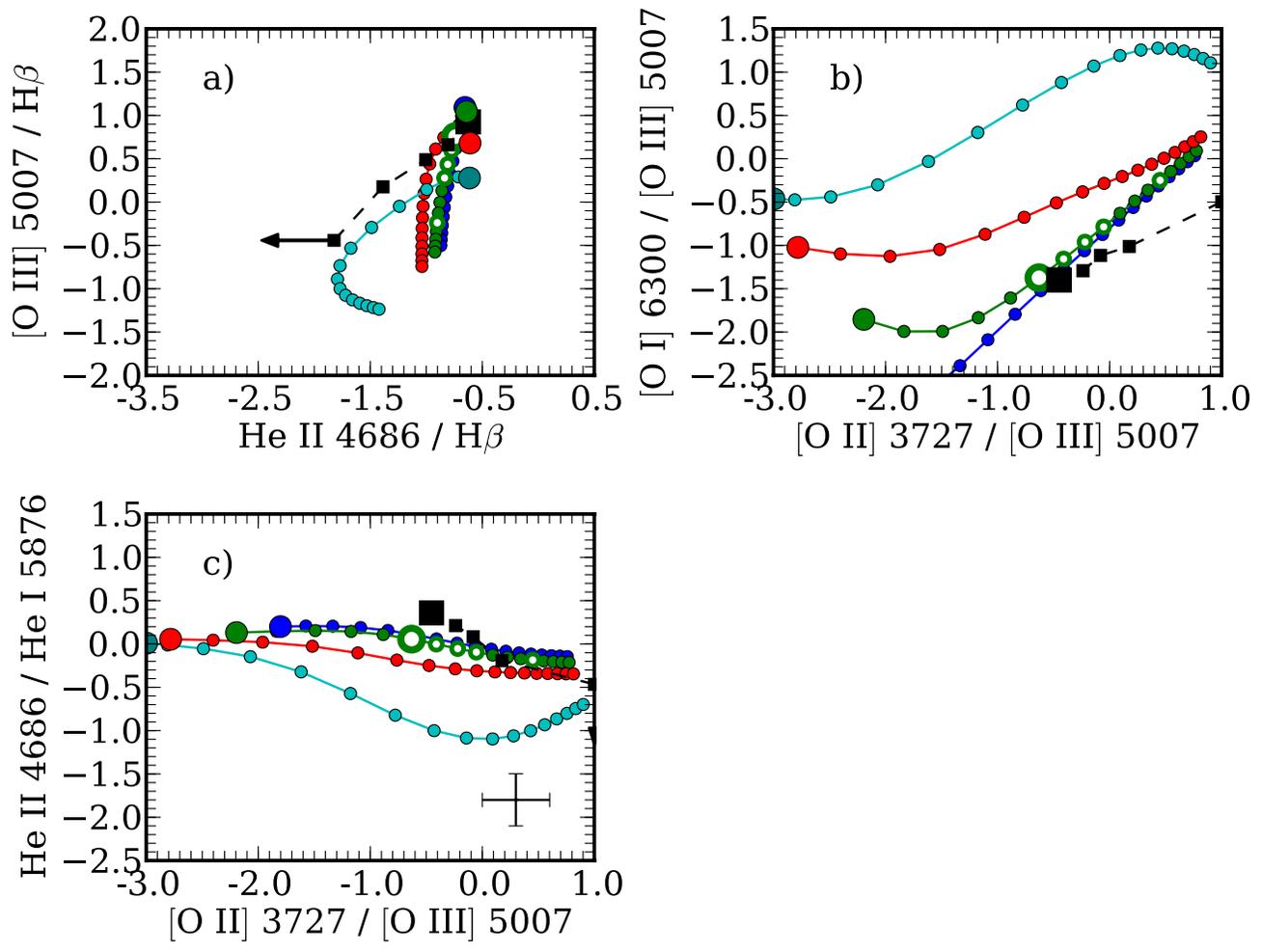

**Figure 11.** Line ratio diagrams that constrain the SED, with lines and symbols the same as Fig. 10. The extreme AGN subsets (large symbols) are successfully fitted in essentially every diagram. Our model reproduces the entire AGN sequence for all but He II λ4686/ Hβ.

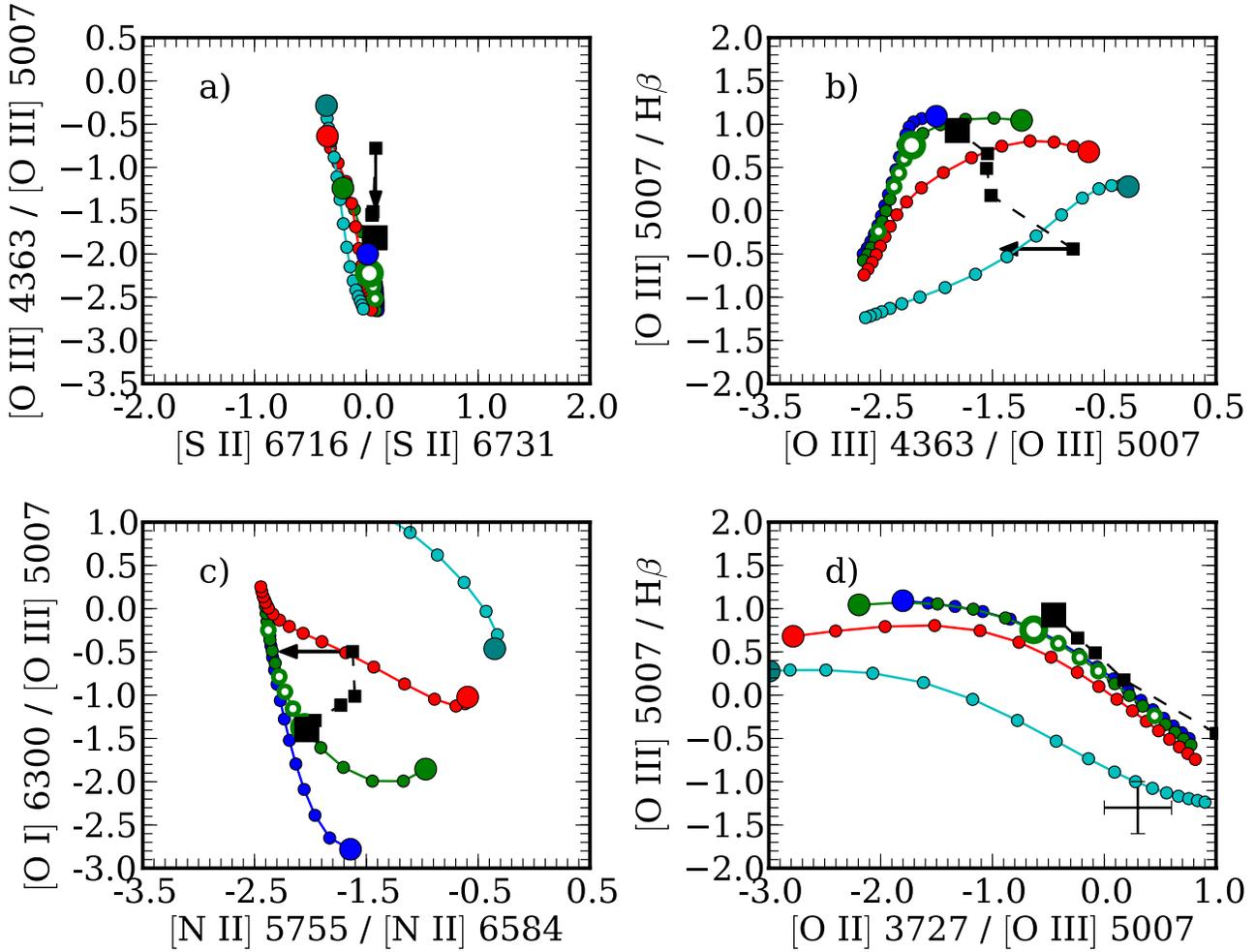

**Figure 12.** Line ratio diagrams that constrain the physical conditions, with lines and symbols the same as Fig. 10. The excellent agreement between our model and the [S II] λ6716 / [S II] λ6731 ratio over the entire range of ionization shows that low density regions are correctly predicted by our model. The [O II] λ3727 / [O III] λ5007 ratio, sensitive to ionization parameter, fits the observations at all but the lowest ionization points. However, the predicted [O III] λ4363 / [O III] λ5008 and [N II] λ5755 / [N II] λ6584 temperature-indicator ratios fail to match the observations except for the highest ionization cases.

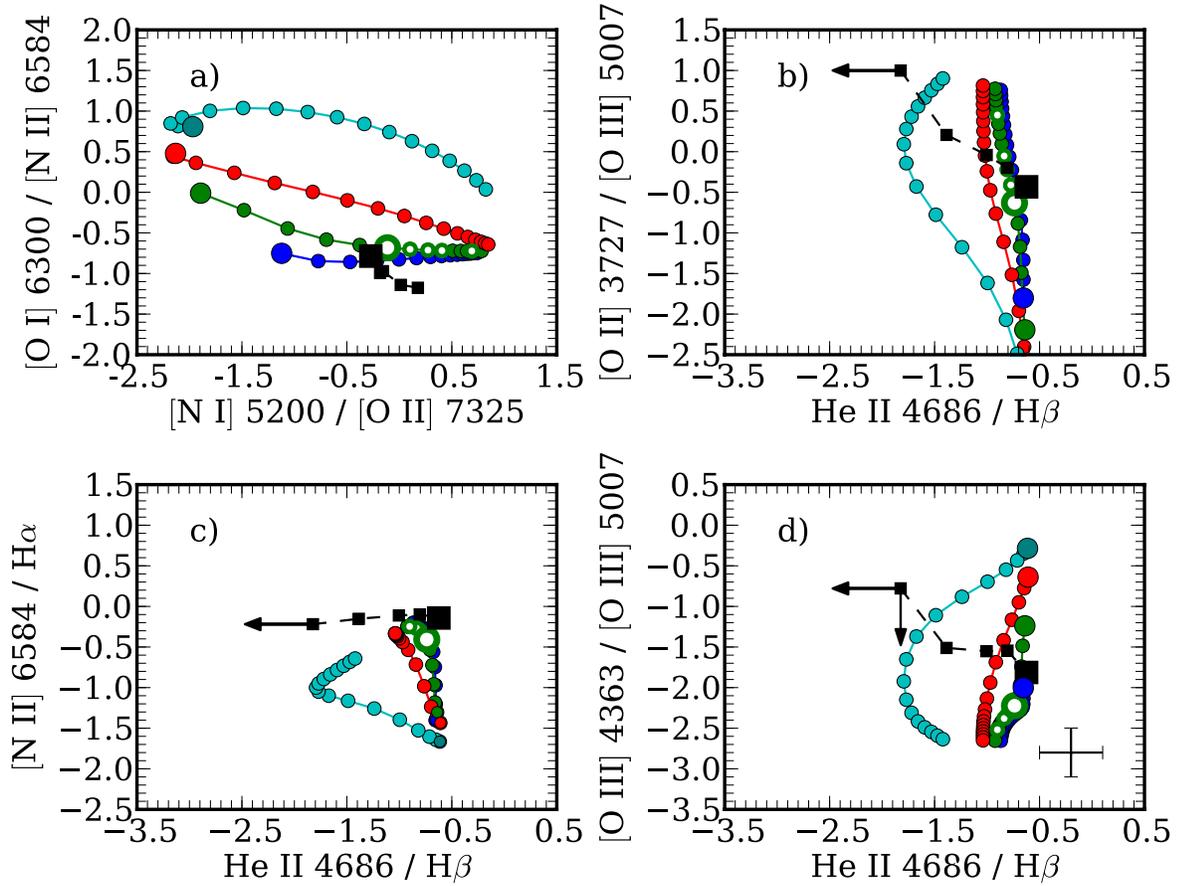

**Figure 13.** Line ratio diagrams sensitive to radiation pressure from grains taken from G04b, with lines and symbols the same as Fig. 10. These dust-free models successfully fit all of the high to moderate ionization observations except for [O III] λ4363 / [O III] λ5008, which also failed in G04b.

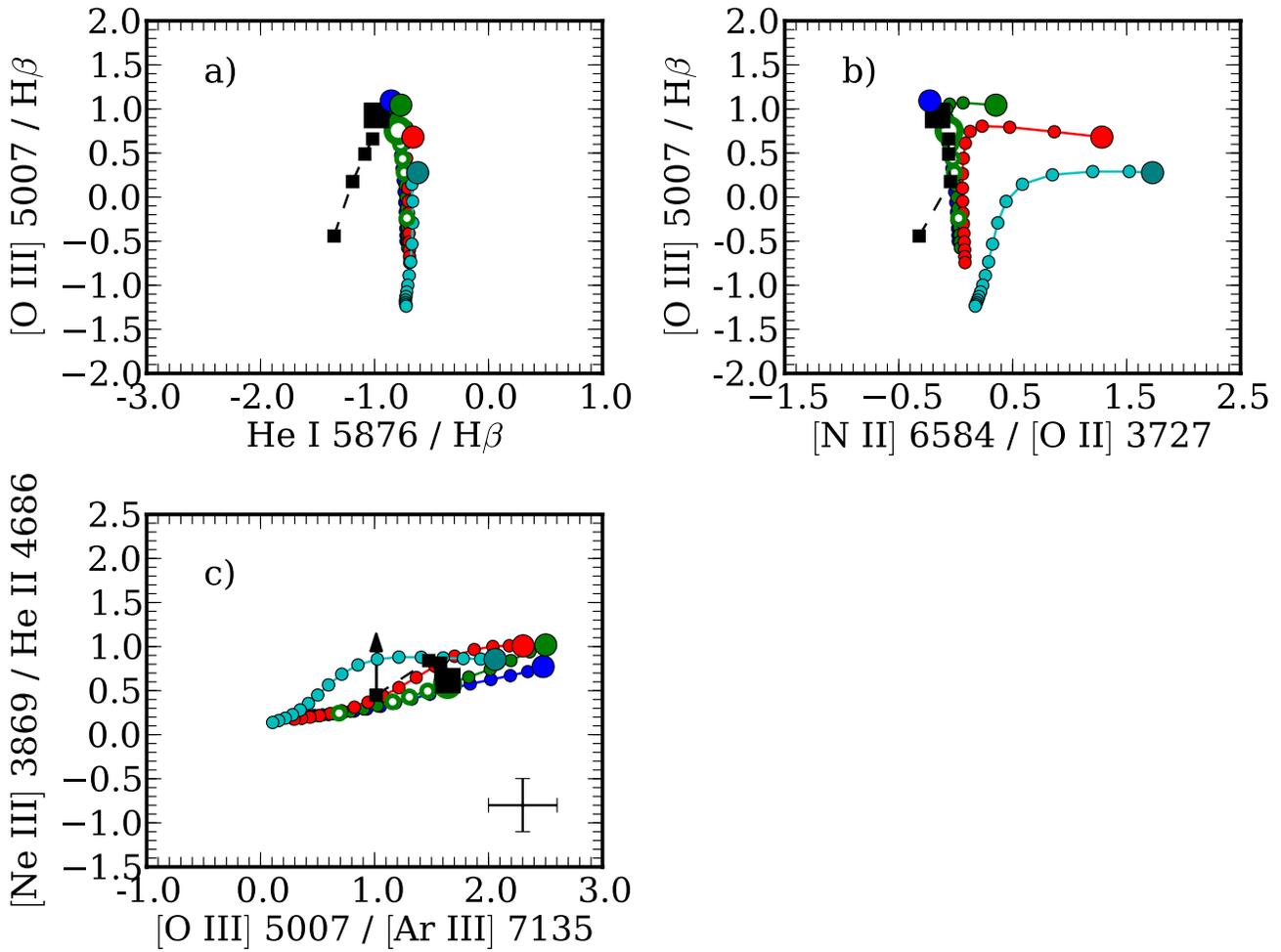

**Figure 14.** Line ratio diagrams that constrain abundances, with lines and symbols the same as Fig. 10. The best abundance sensitive ratio, [N II] λ6584 / [O II] λ3727, was used to optimize our abundance set and fits all but the lowest ionization subset. The models and observations agree to within the factor two error bars for the high through moderate ionization subsets.

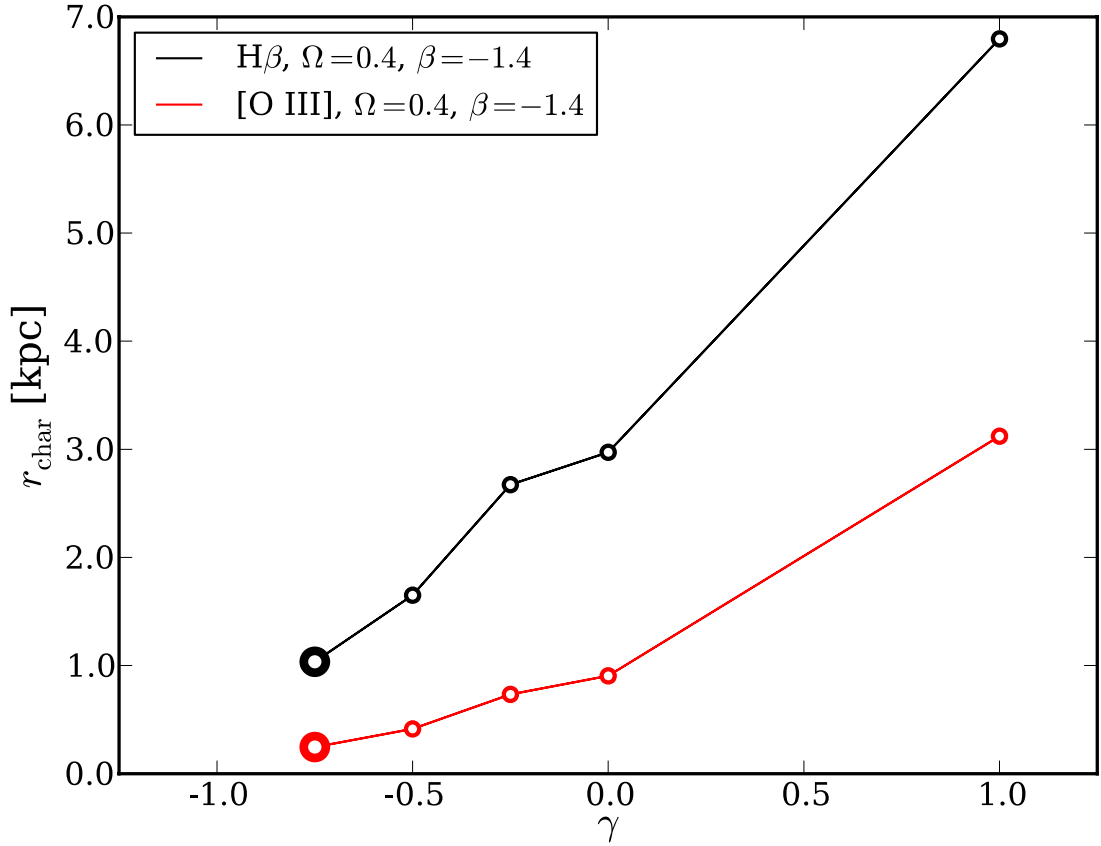

**Figure 15.** Effective NLR sizes for dust-free models that fit the AGN sequence subsets, assuming a covering factor of $\Omega = 0.4$. The LOC density concentration parameter $\beta$ is constant. An LOC model with $\gamma = -0.75$ (indicated by the largest circles) fits the spectrum of the highest ionization AGN subset, a41, while the LOC models with higher $\gamma$ values, indicated by the subsequent smaller circles, fit the lower-ionization observed subsets. Lower ionization AGN are larger in size than their higher ionization counterparts.

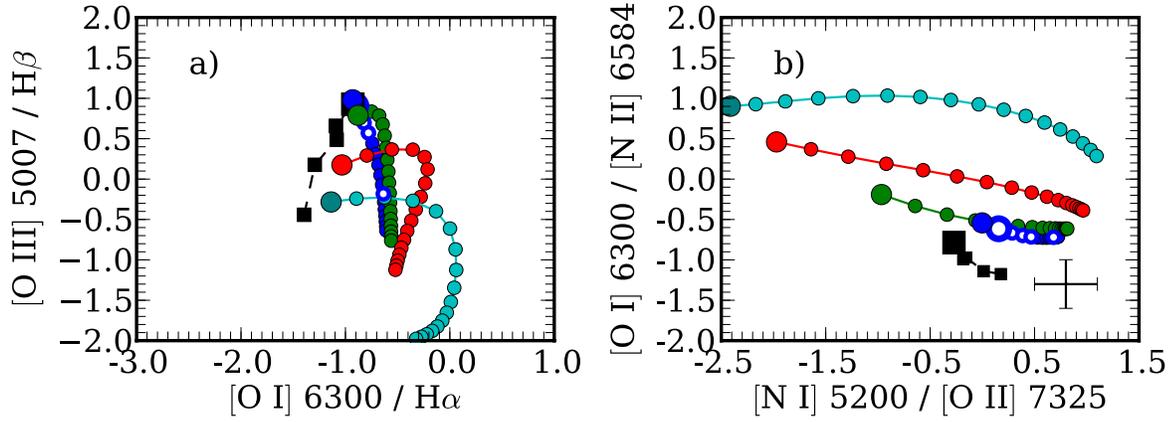

**Figure 16.** Line ratio diagrams for our dusty NLR models with lines and symbols the same as Fig. 10. The AGN observations of some key line ratios, which are matched by our dust-free models, are not fit as well by the dusty models shown here. The hollow marker indicates our best set of free parameters in the dusty case.

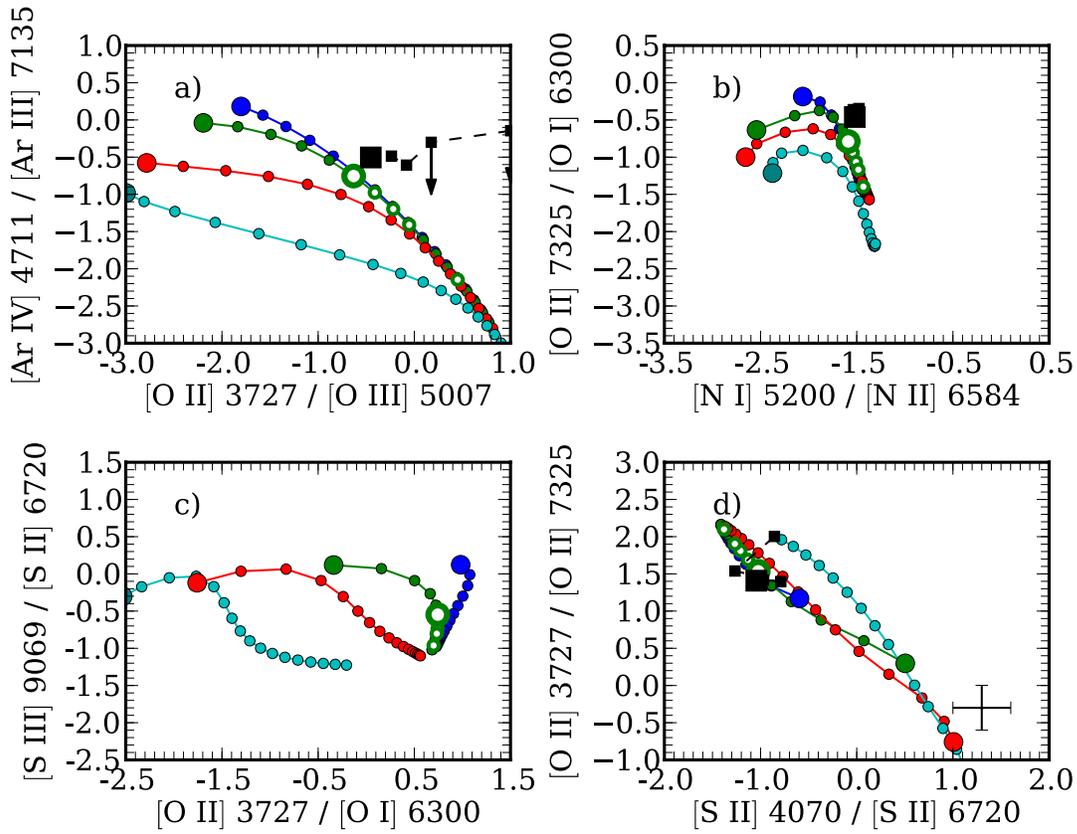

**Figure A1.** Weaker diagnostic diagrams constraining the SED and density with lines and symbols the same as Fig. 10. Our model reproduces the high ionization observations for the two SED diagrams (top panels) except for [O II] λ7325 / [O I] λ6300 which very nearly with a factor of 2. Panel (c) is included for other studies that might include [S III] λ9069. High density regions, probed in Panel (d), also agree with the high-moderate ionization observations.

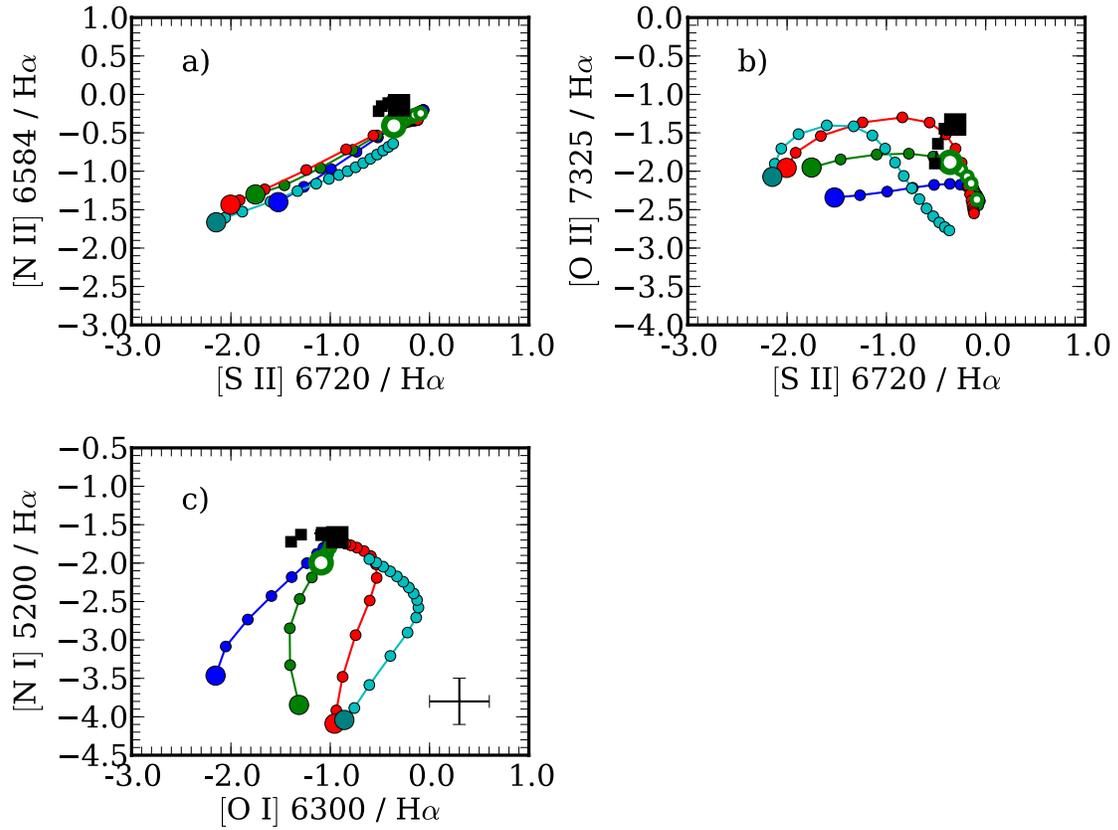

**Figure A2.** Relatively degenerate line ratio diagrams that constrain abundances with lines and symbols in the same as Fig. 10. Our models match the high-moderate ionization subsets for all line ratios, and a few lower ionization subsets, in all cases except for the [O II] λ7325/ Hα ratio, which agrees with the highest ionization subset to within a factor of three.